\documentclass[a4paper, 11pt]{article}
\pdfoutput=1  
\usepackage{jheppub} 

\usepackage{amsmath,amsfonts,amsbsy,amssymb,array,accents,dsfont}
\usepackage{enumerate,array,latexsym,graphicx,mathrsfs,verbatim,psfrag}
\usepackage{enumerate}
\usepackage{graphicx} 
\usepackage[usenames]{color}

\usepackage{bm}
\usepackage{amssymb}
\usepackage{amsmath}
\usepackage{cancel}
\usepackage{epigraph}
\usepackage{verbatim}

\newcommand{\captionfonts}{\small}

\makeatletter  
\long\def\@makecaption#1#2{%
  \vskip\abovecaptionskip
  \sbox\@tempboxa{{\captionfonts #1: #2}}%
 \ifdim \wd\@tempboxa >\hsize
    {\captionfonts #1: #2\par}
  \else
    \hbox to\hsize{\hfil\box\@tempboxa\hfil}%
  \fi
  \vskip\belowcaptionskip}
\makeatother   

\usepackage{fancyhdr}
\usepackage{datetime}

\usepackage[all]{hypcap}     

\makeatletter
\renewcommand\section{\@startsection {section}{1}{\z@}%
                                   {-3.5ex \@plus -1ex \@minus -.2ex}
                                   {2.3ex \@plus.2ex}%
                                   {\normalfont\Large\bfseries}}
\renewcommand\subsection{\@startsection{subsection}{2}{\z@}%
                                     {-3.25ex\@plus -1ex \@minus -.2ex}%
                                     {1.5ex \@plus .2ex}%
                                     {\normalfont\bfseries}}
\renewcommand\subsubsection{\@startsection{subsubsection}{3}{\z@}%
                                     {-2.5ex\@plus -1ex \@minus -.2ex}%
                                     {1.25ex \@plus .2ex}%
                                     {\normalfont\textit}}
\makeatother

\DeclareMathSymbol{\medhatsym}{\mathord}{largesymbols}{"62} 

\DeclareMathSymbol{\medtildesym}{\mathord}{largesymbols}{"65}

\DeclareMathAlphabet{\mathpzc}{OT1}{pzc}{m}{it}

\newcommand{\eq}[1]{\eqref{#1}}

\def\({\left(}
\def\){\right)}
\def\[{\left[}
\def\]{\right]}

\def\k{\ensuremath{\mathsf{k}}}

\newcommand{\qp}{\ensuremath{Q_{p}}}

\def\barray{\begin{array}}
\def\earray{\end{array}}
\def\be{\begin{equation}}
\def\ee{\end{equation}}
\def\bea{\begin{eqnarray}}
\def\eea{\end{eqnarray}}
\def\bal{\begin{align}}
\def\eal{\end{align}}

\definecolor{cardinal}{rgb}{0.6,0,0}
\definecolor{darkgreen}{rgb}{0,0.4,0}
\definecolor{green}{rgb}{0,0.4,0}
\definecolor{golden}{rgb}{0.92, 0.7, 0}
\definecolor{midnight}{rgb}{0, 0, 0.5}
\definecolor{darkblue}{rgb}{0, 0, 0.7}

\numberwithin{equation}{section}

\mathchardef\mhyphen="2D

 \def\cB{\mathcal {B}}

\def\cM{\mathcal {M}}

\def\one{{\hbox{\kern+.5mm 1\kern-.8mm l}}}
\def\zero{{\hbox{0\kern-1.5mm 0}}}
\renewcommand{\k}{\mathsf{k}}

\newcommand{\ket}[1]{{\,| {#1} \rangle}}

\def\^{\wedge}

\usepackage[nottoc]{tocbibind}

\title{\boldmath Shockwaves in black hole microstate geometries}
\author{Bidisha Chakrabarty,}
\author{Sami Rawash,}
\author{David Turton}

\affiliation{University of Southampton,
University Road,
Southampton 
SO17 1BJ,
United Kingdom}

\emailAdd{b.chakrabarty@soton.ac.uk}
\emailAdd{s.rawash@soton.ac.uk} 
\emailAdd{d.j.turton@soton.ac.uk} 

\abstract{Gravitational solutions involving shockwaves have attracted significant recent interest in the context of black holes and quantum chaos. Certain classes of supersymmetric two-charge black hole microstates are described by supergravity solutions containing shockwaves, that are horizonless and smooth away from the shockwave.
These configurations have been used to describe how black hole microstates absorb and scramble perturbations.
In this paper we construct the first family of asymptotically flat supersymmetric three-charge microstate solutions that contain shockwaves.
We identify a family of holographically dual states of the D1-D5 CFT and show that these pass a set of tests, including a precision holographic test. We find precise agreement between gravity and CFT. Our results may prove useful for constructing more general families of black hole microstate solutions.}

\begin{document}
\maketitle
\raggedbottom


\section{Introduction}

Deriving a consistent quantum description of black holes remains a major open problem of fundamental theoretical physics. One of the sharpest obstructions to such a description is the black hole information paradox~\cite{Hawking:1976ra}, which remains a topic of significant current interest. String Theory offers the prospect of accounting for black hole entropy~\cite{Strominger:1996sh}, resolving black hole singularities, and providing a consistent description of black hole evaporation. However much is currently not understood.

Large families of black hole microstates are known to be explicitly describable in String Theory~\cite{Lunin:2001jy,Mathur:2005zp,Bena:2007kg,Skenderis:2008qn}. These results suggest that quantum gravity effects are important on the scale of the black hole horizon, due to the size of the underlying quantum bound state.
In particular, several families of microstates of supersymmetric D1-D5-P black holes are described by smooth horizonless supergravity solutions. 
The state-of-the-art such solutions are known as superstrata, see e.g.~\cite{Bena:2015bea,Bena:2016agb,Bena:2016ypk,Bena:2017geu,Bena:2017xbt,Bena:2017upb,Ceplak:2018pws,Heidmann:2019zws,Heidmann:2019xrd}. The proposed holographic description of superstrata has passed precision tests~\cite{Giusto:2015dfa,Giusto:2019qig,Rawash:2021pik}. The most recently constructed solutions in this programme include both supersymmetric and non-supersymmetric solutions~\cite{Ganchev:2021pgs,Ganchev:2021iwy,Ganchev:2021ewa}.

Gravitational solutions containing shockwaves describe the backreaction of massless point particles~\cite{Aichelburg:1970dh}. 
There has been significant recent interest in such solutions in the context of the behaviour of black holes and quantum chaos. Shockwave collisions on black hole backgrounds probe the absorptive nature of the horizon,  providing insight into the chaotic behaviour of out-of-time-ordered correlators (OTOCs) in the holographically dual CFT~\cite{Shenker:2013pqa,Shenker:2013yza}. This work led to a proposed bound on such chaotic behaviour~\cite{Maldacena:2015waa}, which may be thought of as a refinement of the conjecture that black holes are the fastest scramblers in Nature~\cite{Sekino:2008he}.

Solutions containing shockwaves have also appeared in the context of two-charge black hole microstates. By considering a uniform distribution of high-energy massless point particles, one can obtain a stationary gravitational solution with a shockwave~\cite{Lunin:2002fw,Lunin:2002bj}.
These solutions are deformations of smooth circular supertubes~\cite{Balasubramanian:2000rt,Maldacena:2000dr}, where the shockwave is in the core of the solution. The shockwave describes the backreaction of the high-energy massless quanta, the details of which are not resolved by supergravity. Solutions containing shockwaves can be obtained by a coarse-graining limit of the profile functions that parameterize the general family of two-charge solutions~\cite{Lunin:2001jy,Lunin:2002iz,Taylor:2005db,Kanitscheider:2007wq}.

In recent years there have been several studies of perturbations of microstate geometries. 
Focusing on two-charge circular supertubes and the three-charge spectral flowed supertubes constructed in~\cite{Lunin:2004uu,Giusto:2004id,Giusto:2004ip,Jejjala:2005yu} and studied in \cite{Giusto:2012yz,Chakrabarty:2015foa,Martinec:2017ztd,Martinec:2018nco,Martinec:2019wzw,Martinec:2020gkv,Bufalini:2021ndn}, a classical perturbation analysis was performed in~\cite{Eperon:2016cdd}.
These solutions have a surface of infinite redshift known as an evanescent ergosurface, around which there are stably trapped null geodesics with associated long-lived quasinormal modes. A heuristic argument was presented that a probe massive particle, coupled to supergravity fields, will minimize its energy by approaching a null geodesic at the evanescent ergosurface. The local energy of such a probe would then be large, indicating a potential non-linear classical instability associated with its backreaction. For related work, see~\cite{Keir:2018hnv}.

For two-charge supertubes, it was later argued that the solutions involving shockwaves of~\cite{Lunin:2002fw,Lunin:2002bj} should describe the backreaction of such probes, and that the overall physical process is an evolution from less typical to more typical microstates~\cite{Marolf:2016nwu}. In the solutions of~\cite{Lunin:2002bj} the shockwave is located at the evanescent ergosurface and so the solutions with shockwaves also describe two-charge Ramond-Ramond (RR) ground states.  More recent work has refined this interpretation in the context of scrambling and the resulting motion on the moduli space of RR ground states~\cite{Bena:2018mpb,Martinec:2020gkv}. These microscopic perspectives indicate that the evolution to more typical states, which requires the bound state to shed angular momentum, is constrained by the energy supplied by the perturbation.

Perturbations of three-charge solutions have also been recently investigated. For three-charge solutions, the long-lived quasinormal modes of spectral flowed supertubes can also be derived from the holographically dual CFT~\cite{Chakrabarty:2019ujg}. Furthermore, one can investigate scrambling and chaos in superstrata. It has been found that extremal black holes and their microstates exhibit a slower scrambling than that seen in non-extremal black holes~\cite{Craps:2020ahu}. This slow scrambling can also be seen in the dual two-dimensional CFT~\cite{Craps:2021bmz}.
Relatedly, tidal forces have been computed in superstratum solutions~\cite{Bena:2020iyw}, and analyzed in the holographically dual CFT~\cite{Guo:2021gqd}. Chaotic behaviour has also been observed at the rim of black hole and microstate geometry shadows~\cite{Bianchi:2020des}.

In this paper we construct the first three-charge black hole microstate solutions that contain shockwaves in their core regions. For our seed solutions we take the three-charge BPS fractional spectral flowed supertubes of~\cite{Giusto:2012yz}.
We begin in the two-charge limit, in which the fractional spectral flowed supertubes reduce to circular supertubes. We then consider the deformations of these solutions that contain shockwaves~\cite{Lunin:2002bj}, in the AdS limit. We perform spacetime spectral flow to obtain the shockwave deformation of the fractional spectral flowed supertubes in the AdS limit. We then use the multi-center formalism developed in~\cite{Bena:2004de,Gauntlett:2004qy,Giusto:2004kj,Bena:2005va,Berglund:2005vb} to extend these solutions to new asymptotically flat BPS solutions.

Apart from the shockwave singularity, our solutions are otherwise smooth (up to possible orbifold singularities) and free of closed timelike curves. 
The shockwave is a coarse-grained description of the backreaction of a set of high-energy quantum or set of quanta; for instance we know the total energy of the system, but not how this is distributed among the massless particles~\cite{Lunin:2002bj}. 
This means that our solutions give an approximate collective description of a family of microstates: the supergravity solutions do not resolve the microscopic details of the shockwave.
This is in contrast to the individual pure coherent states described by smooth solutions. 
Nevertheless these new solutions might provide a useful guide for the  construction of more general smooth microstate geometries describing pure states.

Correspondingly, the dual holographic description is not an individual pure state but instead a family of pure dual CFT states that are approximately described by the same bulk solution at the resolution of supergravity.
We propose a specific family of holographically dual CFT states, and perform tests of our proposal, including a precision holographic test.
As a byproduct of our precision holography analysis, we also refine the proposal of~\cite{Lunin:2002bj} for the CFT states dual to the two-charge supertubes with shockwaves.

This paper is organized as follows. In Section~\ref{sec2} we review the two-charge BPS supertube solutions with shockwaves. In Section~\ref{sec3} 
we construct new three-charge microstate solutions containing shockwaves.
In Section~\ref{sec4} we refine the proposal for the CFT states dual to 
circular supertube solutions with shockwaves, propose a family of CFT states dual to our new solutions, and perform tests of this proposal. We discuss our results in Section~\ref{Concls}.

The appendices describe several details of our work. In Appendix \ref{app:sugra}, we record the form of the D1-D5-P 1/8-BPS solution of type IIB supergravity compactified on $T^4$ that corresponds to six-dimensional minimal supergravity coupled to a single tensor multiplet, and the associated BPS equations. 
In Appendix \ref{charges11}, we compute the conserved charges of the supergravity solutions describing fractionally spectral flowed supertubes with shockwaves. We describe the details of our precision holography computation in Appendix \ref{app:prec-holog}.

\section{Shockwaves in supertube backgrounds}
\label{sec2}

In this section we review the supergravity solution that describes a shockwave in a circular supertube background~\cite{Lunin:2002fw,Lunin:2002bj}, and make a straightforward generalization to introduce an orbifold parameter $k$.

We consider Type IIB string theory compactified on $\cM\times S^1$, where $\cM$ is $T^4$ or K3. We take $T^4$ for concreteness. We consider the $T^4$ to be microscopic and the $S^1$ to be macroscopic. We consider bound states of D1 branes wrapped on $S^1$ and D5 branes wrapped on $S^1 \times T^4$. We work in the supergravity limit, with D1 and D5 supergravity charges $Q_1$ and $Q_5$ respectively. We consider configurations that are invariant on the $T^4$, and mostly work in six dimensions. Furthermore, we work in the truncation that corresponds to minimal 6D supergravity coupled to one tensor multiplet; the corresponding Type IIB ansatz and BPS equations are recorded in Appendix \ref{app:sugra}.

We begin in the AdS$_3\times$S$^3$ decoupling limit, in which the original asymptotic $S^1$, coordinatized by $y$, has become the angular direction of AdS$_3$. We consider the background obtained by taking a $\mathbb{Z}_{k}$ orbifold of the global AdS$_3\times S^3$ vacuum, supported by the self-dual two-form potential $C_2$:
\begin{align}\label{twisted vacuum}
\begin{aligned}
    ds^2_6&\;=\,\sqrt{Q_1Q_5}\left( -\frac{1+k^2r^2}{k^2}dt^2+\frac{k^2}{1+k^2r^2}dr^2+r^2dy^2+d\theta^2+\sin^2\theta d\phi^2+\cos^2\theta d\psi^2\right)\,,\\
   C_2&\;=\,\sqrt{Q_1Q_5}\Big(\cos^2\theta d\phi \wedge d\psi+r^2 dt\wedge dy\Big)\,.
    \end{aligned}
\end{align}
In this limit the dilaton is a fixed scalar, $e^{2\Phi}\,=\,Q_1/Q_5$.
One can deform this background to add a shockwave while preserving supersymmetry~\cite{Lunin:2002fw,Lunin:2002bj}. Let us consider a distribution of massless quanta at the center of AdS ($r=0$) and at $\theta=\frac{\pi}{2}$ on the $S^3$, moving in the $\phi$ direction.   We take the energy of each quantum to be large such that we can treat the quanta as massless point particles, and we consider a uniform distribution of such quanta along the $\phi$ coordinate.

The backreaction of this distribution of quanta can be described by a stationary solution involving an Aichelburg-Sexl type shockwave on the above background. For $k=1$ this solution was constructed in~\cite{Lunin:2002fw} and further studied in~\cite{Lunin:2002bj}. The generalization to $k>1$ is straightforward and is given
in terms of a parameter $q$ with $0\leq q<1$ that parametrises the strength of the shockwave: 
\begin{equation}\label{sw+twisted vacuum}
\begin{aligned}
    ds^2&\,=\,\sqrt{Q_1Q_5}\left[ -\frac{1+k^2r^2}{k^2}dt^2+\frac{k^2}{1+k^2r^2}dr^2+r^2dy^2+d\theta^2+\sin^2\theta d\phi^2+\cos^2\theta d\psi^2 \right. \\
    &\qquad\qquad\qquad\qquad \left.
    {}+q\frac{\big((kr^2+1/k)dt+\sin^2\theta d\phi\big)^2-\big(kr^2dy-\cos^2\theta d\psi\big)^2}{k^2r^2+\cos^2\theta}
    \right]\,,\\
     C_2&\,=\,\sqrt{Q_1Q_5}\bigg[\cos^2\theta\, d\phi \wedge d\psi+r^2 dt\wedge dy\\&\qquad\qquad\qquad-
     \frac{q}{k(k^2 r^2+\cos^2\theta)}\bigg(
     k \sin^2\theta \big(-\cos^2\theta d\phi\wedge d\psi+kr^2 d\phi \wedge dy \big)\\&
     \qquad\qquad\qquad\qquad\qquad\qquad\qquad+(1+k^2 r^2)\big(\cos^2\theta d\psi \wedge dt+kr^2dt \wedge dy\big)
     \bigg)
     \bigg]\,.
    \end{aligned}
\end{equation}
Near the locus $(r=0,\theta=\pi/2)$, the metric is approximately
\begin{equation}\label{singularity sw}
    ds^2\simeq \sqrt{Q_1Q_5}\Big[-\frac{dt^2}{k^2}+k^2dr^2+r^2 dy^2+d\theta^2+\cos^2\theta d\psi^2 +d\phi^2+\frac{q}{k^2r^2+\cos^2\theta}\Big(\frac{dt}{k}+d\phi\Big)^2\Big]
\end{equation}
which has a shockwave singularity at $(r=0,\theta=\pi/2)$. For $k=1$ this is an Aichelburg-Sexl-type shockwave generalized to 5+1 dimensions and smeared along the shockwave locus~\cite{Lunin:2002fw}. For $k>1$ the shockwave singularity is located at the $\mathbb{Z}_k$ orbifold singularity of the solution in Eq.~\eqref{twisted vacuum}. 

Upon spectral flow to the Ramond-Ramond (RR) sector, this solution gives an approximate description of a family of RR ground states of the dual CFT, as we shall review in Section~\ref{sec:CFT supertube+sw}.
The relevant spacetime (fractional) spectral flow coordinate transformation is as follows:
\begin{equation}\label{supertube SF}
\phi\rightarrow\phi+ \frac{t}{k}\,,\qquad \psi\rightarrow\psi+\frac{y}{k}\,.
\end{equation}
The result of this coordinate transformation is a 1/4-BPS two-charge microstate solution describing the backreaction of a shockwave on a circular supertube geometry, still so far in the AdS$_3$ decoupling limit.

We now extend the AdS solution to an asymptotically flat ($\mathbb{R}^{1,4}\times S^1$)
solution. For $k=1$ this was done  in~\cite{Lunin:2002bj} and again we make the straightforward generalization to $k>1$. To do so we introduce the scale $R_y$ that will become the asymptotic radius of the $y$ circle, and a scale $a$ defined in the following equation. We define dimensionful coordinates via the rescaling
\begin{equation}
    r\rightarrow \frac ra\,,\qquad t\rightarrow t R_y\,, \qquad y\rightarrow y R_y\,, \qquad a^2 \;=\; \frac{Q_1 Q_5}{k^2 R_y^2}.
\end{equation}
The extension of this solution to an asymptotically flat one was obtained, for $k=1$, in~\cite{Lunin:2002bj}, generalizing the two-charge circular supertube solutions (without shockwaves) of~\cite{Balasubramanian:2000rt,Maldacena:2000dr}.
The straightforward generalization to arbitrary $k$ gives the following solution:
\begin{eqnarray}
ds^2&\;\!=\;\!&-\frac{1}{\bar h_{(0)}}(dt^2-dy^2)+\bar h_{(0)} \bar f_{(0)}\left(d\theta^2+\frac{k^2 d\bar r^2}{ k^2 {\bar r}^2+\bar a^2}\right)-\xi \frac{2 a \sqrt{Q_1Q_5}}{k\bar h_{(0)} \bar f_{(0)}}(\cos^2 \theta \,dy\,d\psi+\sin^2\theta\, dt\,d\phi)
\nonumber\\[0.5mm]
&&{}+\bar h_{(0)}\Big[\Big(\bar r^2+\xi \frac{\bar a^2 Q_1Q_5\cos^2\theta}{k^2\bar h_{(0)}^2 \bar f_{(0)}^2}\Big)\cos^2\theta\,d\psi^2+\Big(\bar r^2+\frac{\bar a^2}{k^2}-\xi\frac{\bar a^2 Q_1Q_5 \sin^2\theta}{k^2\bar h_{(0)}^2 \bar f_{(0)}^2}\Big)\sin^2 \theta \,d\phi^2\Big]\,,\nonumber\\[1.5mm]
C_2&\;\!=\;\!&-\frac{Q_1 dt\wedge dy}{\bar f_{(0)}\bar h_{1(0)}}-\frac{a\,\xi \sqrt{Q_1Q_5}}{k\bar f_{(0)}\bar h_{1(0)}}\Big(\cos^2\theta dt\wedge d\phi+\sin^2\theta dy\wedge
d\phi\Big)\label{superstube+sw}\\[0.5mm]
&&{}+\Big(\frac{\bar a^2 q\, Q_1Q_5 \sin^2\theta}{k^2\bar f_{(0)}^2\bar h_{1(0)}}+
\frac{Q_5 (k^2 Q_1+k^2 \bar f_{(0)}+\bar a^2 \sin^2\theta)}{k^2\bar f_{(0)}\bar h_{1(0)}}\Big)\cos^2\theta d\phi\wedge d\psi\,,
\nonumber\\[0.5mm]
e^{2\Phi}&\;\!=\;\!&\frac{\bar h_{1(0)}}{\bar h_{5(0)}}\,,\nonumber
\end{eqnarray}
where $\xi=1-q$ parametrises the strength of the shockwave, and where
\begin{equation}
    \begin{aligned}
        \bar r&\,=\,\sqrt\xi r\,,\qquad\qquad \bar a\,=\,\sqrt\xi a\,,\qquad\qquad \bar f_{(0)}\,=\,\xi(r^2+\frac{a^2}{k^2}\cos^2\theta)\,,\\
       \bar h_{1(0)}&\,=\,1+\frac{Q_1}{\bar f_{(0)}}\,,\qquad
        \bar h_{5(0)}\,=\,1+\frac{Q_5}{\bar f_{(0)}}\,,\qquad \bar h_{(0)}\,=\,\sqrt{ \bar h_{1(0)} \bar h_{5(0)}}\;.
    \end{aligned}
\end{equation}
The subscript $(0)$ denotes supertube quantities and we use it to distinguish the above functions from those that characterize the new solutions that we will report in the next section.

\section{Shockwaves in fractionally spectral flowed supertubes}
\label{sec3}
In this section we first review the three-charge, 1/8-BPS, fractionally spectral flowed supertube solutions constructed and studied in~\cite{Lunin:2004uu,Giusto:2004id,Giusto:2004ip,Jejjala:2005yu,Giusto:2012yz}, as well as their decomposition into two-center solutions of the multi-center formalism of~\cite{Bena:2004de,Gauntlett:2004qy,Giusto:2004kj,Bena:2005va,Berglund:2005vb}. We then proceed to construct a novel family of BPS solutions involving shockwave deformations of these solutions.

\subsection{Fractionally spectral flowed circular supertubes}
\label{sec31}
Fractionally spectral flowed circular supertubes are a family of 1/8-BPS microstates of the D1-D5-P system. In addition to their D1 and D5 charges, they carry momentum charge along $y$ that we denote by $\qp$. 
The solutions take the form~\cite{Giusto:2004id,Giusto:2004ip,Giusto:2012yz}
\begin{equation}\label{GLMT}
\begin{aligned}
    ds^2\,=\,&-\frac{1}{h}(dt^2-dy^2)+\frac{\qp}{h f}(dt-dy)^2+hf\left(\frac{dr^2}{r^2+a^2(\gamma_1+\gamma_2)^2\eta}+d\theta^2\right)\\
    &+h\Big(r^2+a^2\gamma_1(\gamma_1+\gamma_2)\eta -\frac{Q_1Q_5a^2(\gamma_1^2-\gamma_2^2)\eta\cos^2\theta}{h^2f^2}\Big)\cos^2\theta d\psi^2\\
    &+h\Big(r^2+a^2\gamma_1(\gamma_1+\gamma_2)\eta +\frac{Q_1Q_5a^2(\gamma_1^2-\gamma_2^2)\eta\sin^2\theta}{h^2f^2}\Big)\sin^2\theta d\phi^2
    \\
    &+\frac{\qp\, a^2(\gamma_1+\gamma_2)^2\eta^2}{hf}(\cos^2\theta d\psi+\sin^2\theta d\phi)^2\\
    &-\frac{2\sqrt{Q_1Q_5}\,a}{hf}(\gamma_1\cos^2\theta d\psi+\gamma_2\sin^2\theta d\phi)(dt-dy)\\
    &-\frac{2\sqrt{Q_1Q_5}\,a(\gamma_1+\gamma_2)\eta}{hf}(\cos^2\theta d\psi+\sin^2\theta d\phi)dy\,,\\[2mm]
    C_2\,=\,&-\frac{\sqrt{Q_1Q_5}\,a\cos^2\theta}{H_1f}(\gamma_2 dt+\gamma_1 dy)\wedge d\psi-\frac{\sqrt{Q_1Q_5}\,a\sin^2\theta}{H_1f}(\gamma_1 dt+\gamma_2 dy)\wedge d\phi\\
    &+\frac{(\gamma_1+\gamma_2)\,a\,\eta \,\qp }{\sqrt{Q_1Q_5}H_1f}(Q_1 dt+Q_5 dy)\^(\cos^2\theta d\psi+\sin^2\theta d\phi)\\
    &-\frac{Q_1}{H_1f}dt\^dy-\frac{Q_5\cos^2\theta}{H_1f}(r^2+\gamma_2(\gamma_1+\gamma_2)\eta +Q_1)d\psi\^d\phi\,,
    \end{aligned}
\end{equation}
    \begin{equation}\label{GLMTdil}
\begin{aligned}
    e^{2\Phi}\,=\,&\frac{H_1}{H_5}\,,
    \end{aligned}
\end{equation}
where the parameters $\gamma_1,\gamma_2$ are determined by integer parameters $s$ and $k$ through
\begin{equation} \label{eq:gsk}
\gamma_1\,=\,-\frac{s}{k}\,,\qquad\gamma_2\,=\,\frac{s+1}{k}    \,,
\end{equation}
and where
\begin{align}
    \begin{aligned} \label{glmt-extras}
   a&\,=\,\frac{\sqrt{Q_1Q_5}}{R}\,,\qquad \qp \,=\,a^2\gamma_1\gamma_2\,,\qquad \eta\,=\,\frac{Q_1Q_5}{Q_1Q_5+Q_1\qp +Q_5\qp }\,,\\
   f&\,=\,r^2+a^2(\gamma_1+\gamma_2)\eta(\gamma_1\sin^2\theta+\gamma_2\cos^2\theta)\,,
   \\
   H_1&\,=\,1+\frac{Q_1}{f}\,,\qquad H_5\,=\,1+\frac{Q_5}{f}\,,\qquad h\,=\,\sqrt{H_1H_5}\,.
    \end{aligned}
\end{align}
In the limit $s\rightarrow 0$ these solutions reduce to the two-charge circular supertube solution of~\cite{Balasubramanian:2000rt,Maldacena:2000dr}.

One can decompose these solutions into the form of the general BPS ansatz for such solutions~\cite{Gutowski:2003rg,Giusto:2013rxa}; this was done in~\cite{Giusto:2004kj,Berglund:2005vb} (see also \cite{Bena:2005va}).
We will use this formalism to construct our solutions, so we now briefly review it and introduce appropriate notation.

The relevant supergravity ansatz is recorded in Appendix~\ref{app:sugra}.  
Supersymmetry and the $U(1)\times U(1)$ isometries along $\phi$ and $\psi$ imply that the base metric   $ds^2_4(\mathcal{B})$ introduced in the second line of \eqref{ansatzSummary} is of Gibbons-Hawking form,
\begin{equation}\label{ds4 GH}
ds_4^2(\mathcal{B})\,=\,V^{-1}(d\varphi_1+A)^2+Vds^2_3   \;, 
\end{equation}
where $ds^2_3$ is the flat metric on $\mathbb{R}^3$, $V$ is a harmonic function on $\mathbb{R}^3$, $A$ is a one-form related to $V$ via $\;\star_3\,dA=dV$, and where $\varphi_1=\phi-\psi$. On such a base metric, solutions can be constructed in terms of a set of multi-center harmonic functions on $\mathbb{R}^3$~\cite{Bena:2004de,Gauntlett:2004qy}, which have poles (centers) at the same points $x^i$ on $\mathbb{R}^3$ (here $I=1,2,3$):
\begin{equation}\label{harmonic functions GH}
    V\,=\,\sum_i\frac{q^{(i)}}{|x-x^i|}\,,\quad~   K_I\,=\,\sum_i\frac{d_I^{(i)}}{|x-x^i|}\,,\quad~   L_I\,=\,\ell_I+\sum_i\frac{Q_I^{(i)}}{|x-x^i|}\,,\quad~   M\,=\,\sum_i\frac{m^{(i)}}{|x-x^i|}\;.
\end{equation}
The relations between these harmonic functions and the quantities $Z_I$, $\Theta^I$, $\beta$, $\omega$ and $\mathcal{F}$ that appear in the BPS ansatz in Appendix~\ref{app:sugra} are given by (see e.g.~\cite{Giusto:2012yz,Bena:2007kg,Bena:2017geu})
\begin{equation}\label{GH to 6dim ans 1}
\begin{aligned}
    Z_I&\,=\,L_I+\frac{1}{2}C_{IJK}\frac{K_JK_K}{V}\,,\qquad \Theta_I\,=\,dB_I\,,\qquad B_I\,=\,\frac{K_I}{V}(d\varphi_1+A)+\xi_I\,,\\
\mathcal{F}&\,=\,-Z_3\,,\qquad \beta\,=\,\frac{K_3}{V}(d\varphi_1+A)+\xi_3\,,\qquad \omega\,=\,\mu(d\varphi_1+A)+\bar \omega\,,
\end{aligned}
\end{equation}
where
\begin{equation}\label{GH to 6dim ans 2}
    \begin{aligned}
    \star_3dK_I&\,=\,-d\xi_I\,,\qquad \mu\,=\,\frac{M}{2}+\frac{K_IL_I}{2V}+\frac{1}{6}C_{IJK}\frac{K_IK_JK_K}{V^2}\,,\\
    \star_3d\bar \omega&\,=\,\frac{1}{2}\Big(VdM-MdV+K_IdL_I-L_IdK_I\Big)\,.
    \end{aligned}
\end{equation}

Asymptotically flat solutions are obtained by setting $\ell_I = 1 ~~\forall\;I$, while in the AdS$_3$ decoupling limit we have instead $\ell_1=\ell_2=0$, $\;\ell_3=1$. Furthermore, in smooth horizonless solutions, the set of coefficients ${q^{(i)},d_I^{(i)},Q_I^{(i)},m^{(i)}}$ in~\eqref{harmonic functions GH} must obey certain constraints~\cite{Bena:2005va,Berglund:2005vb}. Firstly, flat $\mathbb{R}^{1,4}\times$S$^1$ asymptotics and at most local orbifold singularities require that $q^{(i)}\in\mathbb{Z}$ and $\sum_i q^{(i)}=1$.
Next, the coefficients $d^{(i)}_I$ are quantized in terms of integers $k^{(i)}_I$ as (see e.g.~\cite{Giusto:2012yz})
\begin{equation}
    d^{(i)}_1\,=\,\frac{g_s\alpha'}{2R_y}k^{(i)}_1\,,\qquad
    d^{(i)}_2\,=\,\frac{g_s\alpha'^3}{2V_{4}R_y}k^{(i)}_2\,,\qquad
    d^{(i)}_3\,=\,\frac{R_y}{2}k^{(i)}_3\,,
\end{equation}
where the volume of $T^4$ is $(2\pi)^4V_{4}$. 
Regularity of the solution (up to possible orbifold singularities) requires a cancellation of the poles in the harmonic functions~\eqref{harmonic functions GH}: this is ensured if
\begin{equation}\label{GH constraint 3}
    Q^{(i)}_I\,=\,-\frac{|\epsilon_{IJK}|}{2}\frac{d^{(i)}_Jd^{(i)}_K}{q^{(i)}}\,,\qquad m^{(i)}\,=\,\frac{d^{(i)}_1d^{(i)}_2d^{(i)}_3}{(q^{(i)})^2}\,.
\end{equation}
Moreover, absence of CTCs partially constrains the positions of the poles $x^i$:
\begin{equation}\label{GH constraint 4}
    \sum_{j\neq i}\Pi^{(ij)}_1\Pi^{(ij)}_2\Pi^{(ij)}_3\frac{q^{(i)}q^{(j)}}{|x^i-x^j|}\,=\,-\sum_Id^{(i)}_I,\qquad \text{with}\qquad \Pi^{(ij)}_I\,=\,\frac{d^{(i)}_I}{q^{(j)}}-\frac{d^{(j)}_I}{q^{(i)}}\,.
\end{equation}

Fractionally spectral flowed supertubes are two-center solutions~\cite{Giusto:2004kj,Giusto:2012yz}. Indeed they are the most general asymptotically flat such solutions that are regular up to orbifold singularities (which in turn are known to be resolved in the string theory description of these backgrounds~\cite{Martinec:2017ztd,Martinec:2018nco,Martinec:2019wzw,Martinec:2020gkv}).

We introduce spherical polar coordinates centered on the locations of the two centers, $(r_+,\theta_+,\varphi_2)$ and $(r_-,\theta_-,\varphi_2)$, where $\varphi_2=-(\psi+\phi)$. The poles in the harmonic functions~\eqref{harmonic functions GH} are then located at $r_+=0$ and $r_-=0$. The flat $ds^2_3$ base takes the form
\begin{equation}
    ds_3^2\,=\,dr_+^2+r_+^2(d\theta_+^2+\sin^2\theta_+d\varphi_2^2)\,=\,dr_-^2+r_-^2(d\theta_-^2+\sin^2\theta_-d\varphi_2^2)\,,
\end{equation}
where
\begin{equation}
\begin{aligned}
    r_+&\,=\,\frac{r^2+a^2(\gamma_1+\gamma_2)^2\eta\sin^2\theta}{4}\,,\qquad \cos\theta_+\,=\,
   \frac{ r^2\cos 2\theta-a^2(\gamma_1+\gamma_2)^2\eta\sin^2\theta}{r^2+a^2(\gamma_1+\gamma_2)^2\eta\sin^2\theta}\,,\\
      r_-&\,=\,\frac{r^2+a^2(\gamma_1+\gamma_2)^2\eta\cos^2\theta}{4}\,,\qquad \cos\theta_-\,=\,\frac{r^2\cos2\theta+a^2(\gamma_1+\gamma_2)^2\eta\cos^2\theta}{r^2+a^2(\gamma_1+\gamma_2)^2\eta\cos^2\theta}\,.\\
    \end{aligned}
\end{equation}

In our conventions the functions $L_I$ for $I=1,2,3$ correspond to the (electric) D1, D5 and P charges respectively. Writing  $Q_2^{\pm}=Q_5^{\pm},  Q_3^{\pm}=\qp^{\pm},$ the coefficients of the poles in the decomposition of the fractionally spectral flowed solutions \eqref{GLMT} are
\begin{align}\label{coefficients GLMT}
\begin{aligned}
&q^-=-s\,,\qquad q^+=s+1\,,\qquad d_1^-=-d_1^+=Q_5\frac{s(s+1)}{2R_yk}\,,\qquad
d_2^-=-d_2^+=Q_1\frac{s(s+1)}{2R_yk}\,,\\
&d_3^-=-d_3^+=\frac{R_y k}{2}\,,\qquad
Q_1^-=\frac{Q_1 (s+1)}{4}\,,\qquad
Q_1^+=-\frac{s Q_1 }{4}\,,\qquad
Q_5^-=\frac{Q_5 (s+1)}{4}\,,\\
& Q_5^+=-\frac{s Q_5 }{4}\,,\qquad
\qp^-=\frac{Q_1Q_5 s (s+1)^2}{4R_y^2 k^2}\,,\qquad
\qp^+=-\frac{ Q_1 Q_5s^2(1+s) }{4k^2 R_y^2}\,,\\
& m^-=\frac{Q_1Q_5(s+1)^2}{8kR_y}\,,\qquad
m^+=-\frac{Q_1Q_5 s^2}{8kR_y}\,,\qquad \ell_I = 1 ~~~\forall~ I\,.
\end{aligned}
\end{align}
We note that the relations \eqref{GH constraint 3}, \eqref{GH constraint 4} are satisfied.

In the AdS$_3$ decoupling limit, the solution~\eqref{GLMT} is related via a fractional spectral flow large coordinate transformation to the vacuum solution~\eqref{twisted vacuum}. In order to exhibit this, we first take the limit in which the $R_y$ is much larger than the scale set by the $Q_1$ and $Q_5$ charges:
\begin{equation}
    \epsilon\,=\,\frac{(Q_1Q_5)^{1/4}}{R_y} \ll 1 \qquad\Rightarrow\qquad \qp \ll \sqrt{Q_1Q_5}\,,\qquad \eta\simeq 1\,.
\end{equation}
Physically, this regime implies that the geometry~\eqref{GLMT} has an AdS throat whose proper length is large in AdS units (see e.g.~\cite{Mathur:2011gz}). The AdS throat is the region of spacetime where $r \ll \sqrt{Q_1Q_5}$. To take the decoupling limit, we rescale coordinates as
\begin{equation}
    r\rightarrow a\,r\,,\qquad t\rightarrow \frac{t}{R_y}\,, \qquad y\rightarrow \frac{y}{R_y}\,,
\end{equation}
and send $R_y \to \infty$ holding fixed the rescaled dimensionless coordinates $(r,t,y)$ and the charges $Q_1$, $Q_5$. From \eqref{glmt-extras} this sends $a\to 0$, and likewise $\epsilon\to 0$. We then obtain the decoupled metric
\begin{equation}\label{GLMT decoupling limit}
\begin{aligned}
    ds^2\,=\,\sqrt{Q_1Q_5}\Big[& -\frac{1+k^2r^2}{k^2}dt^2+\frac{k^2}{1+k^2r^2}dr^2+r^2dy^2+d\theta^2\\
    &+\sin^2\theta (d\phi-\gamma_2dt-\gamma_1dy)^2+\cos^2\theta (d\psi-\gamma_2dy-\gamma_1dt)^2\Big]\,.
    \end{aligned}
\end{equation}
The fractional spectral flow coordinate transformation
\begin{equation}\label{GLMT SF}
\phi\rightarrow\phi+\gamma_2t+\gamma_1y\,,\qquad \psi\rightarrow\psi+\gamma_1t+\gamma_2y\,,
\end{equation}
maps the geometry in Eq.~\eqref{GLMT decoupling limit} into the $k$-orbifolded global AdS$_3 \times$S$^3$ solution given in Eq.~\eqref{twisted vacuum}.

\subsection{Shockwaves in fractionally spectral flowed supertubes}\label{GLMT+sw}

We now construct three-charge solutions involving shockwaves using a straightforward two-step procedure. In the first step we take the solution involving a shockwave on global AdS~\eqref{sw+twisted vacuum} and apply the inverse of the fractional spectral flow coordinate transformation~\eqref{GLMT SF} to obtain a shockwave deformation of the AdS$_3$ limit of the fractionally spectral flowed circular supertubes. 
For later use we record the resulting metric:
\begin{equation}
\begin{aligned} \label{eq:GLMT+sw-AdS}
    ds^2\,=\,&\sqrt{Q_1Q_5}\bigg[-\frac{(1+k^2 r^2)}{k^2}dt^2+r^2 dy^2+\frac{k^2 dr^2}{1+k^2r^2}+d\theta^2 \\&\hspace{4em}+\cos^2\theta(-\gamma_1dt-\gamma_2dy+d\psi)^2+(-\gamma_2dt-\gamma_1dy+d\phi)^2\sin^2\theta\\
    &\hspace{4em}+\frac{q}{k^2r^2+\cos^2\theta}\bigg(-\big(kr^2dy-(-\gamma_1dt-\gamma_2dy+d\psi)\cos^2\theta\big)^2\\&\hspace{12em}
    +\Big(\frac{1+k^2 r^2}{k}dt+(-\gamma_2dt-\gamma_1 dy+d\phi)\sin^2\theta\Big)^2\bigg)
    \bigg].
    \end{aligned}
\end{equation}

In the second step we extend this solution to an asymptotically flat solution. The method is again straightforward, however the calculation is more involved than the trivial first step. The method is to decompose the solution obtained in the first step into the harmonic functions of the multi-center formalism, and then ``add back the 1'' in the relevant harmonic functions.

To write the decomposition of the solution obtained in the first step, we rescale the location of the two poles of the harmonic functions of the undeformed solution as
\begin{equation}
 r^\pm\;\rightarrow\;\xi \,r^\pm  \,. 
\end{equation}
The coefficients of the two poles are then
\begin{align}\label{coefficients GLMT+sw}
\begin{aligned}
&q^-=-s\,,\qquad q^+=s+1\,,\qquad d_1^-=-d_1^+=Q_5\frac{s(s+1)}{2R_yk}\,,\qquad
d_2^-=-d_2^+=Q_1\frac{s(s+1)}{2R_yk}\,,\\
&d_3^-=-d_3^+=\frac{R_y k}{2}\,,\qquad
Q_1^-=\frac{Q_1 (s+1)}{4}\,,\qquad
Q_1^+=-\frac{s Q_1 }{4}\,,\qquad
Q_5^-=\frac{Q_5 (s+1)}{4},\\
& Q_5^+=-\frac{s Q_5 }{4}\,,\qquad
\qp^-=\frac{Q_1Q_5 s (s^2+2s+\xi)}{4R_y^2 k^2}\,,\qquad
\qp^+=-\frac{ Q_1 Q_5s^2(1+s) }{4k^2 R_y^2}\,,\\
& m^-=\frac{Q_1Q_5(s^2+2s+\xi)}{8kR_y}\,,\qquad
m^+=-\frac{Q_1Q_5 s^2}{8kR_y}, \qquad \ell_1 = \ell_2 = 0 \,, \quad \ell_3=1\,.
\end{aligned}
\end{align}
Having expressed the AdS$_3$ solution in this form, we trivially extend the solution to asymptotically flat space by replacing $\ell_I=1~~\forall\;I$.

To generate the closed-form solution describing a shockwave on the fractional spectral flowed supertube background, we use Eqs.~\eqref{GH to 6dim ans 1} and \eqref{GH to 6dim ans 2} to obtain
\begin{equation}\label{GLMT+sw Q1Q5}
    \begin{aligned}
    ds^2=&\;\frac{\sqrt{\bar h_1 \bar h_5}\bar f dr^2}{b^2+r^2}+\bar f\sqrt{\bar h_1\bar h_5}d\theta^2+\frac{(-dt^2+dy^2)}{\bar h_1\bar h_5}\\
    &+\frac{\cos^2\theta}{\bar f^2\sqrt{\bar h_1\bar h_5}}\bigg[\xi\bar h_1\bar h_5 \,\bar f^2 (r^2 -s\,b^2)+b^2\, Q_1Q_5 \,(2s+1)\,\xi^2 \cos^2\theta\\
    &\hspace{3.5em}    
    -q\Big(b^2\,s\,(-Q_1Q_5+r^2\,(s+1)(Q_1+Q_5))\,\xi +Q_1Q_5\,r^2\Big(\eta\, \xi-\frac{\bar f}{r^2+b^2\cos^2\theta}\Big)\Big)\bigg]d\psi^2
    \\
    &+q\frac{2a\,\sqrt{Q_1Q_5}\,\sin^2\theta(r^2-b^2\,s)(dt-dy)d\phi}{k\bar f\sqrt{\bar h_1\bar h_5}(r^2+b^2\cos^2\theta)}+\frac{a^2\,s(dt-dy)^2(\bar f+s\,(r^2+b^2\sin^2\theta))}{k^2 \bar f \sqrt{\bar h_1\bar h_5}(r^2+b^2\cos^2\theta)}\\
    &+\frac{\sin^2\theta}{\bar f^2\sqrt{\bar h_1\bar h_5}}\Big(\bar h_1\bar h_5 \bar f^2\,(r^2+b^2\,(s+1))\xi-b^2\,Q_1Q_5\,(2s+1)\,\xi\,\sin^2\theta+\frac{q\,b^2 \bar f \,Q_1Q_5\sin^2\theta}{r^2+b^2\cos^2\theta}\Big)d\phi^2
    \\
    &-\frac{2a\,\sqrt{Q_1Q_5}\,\eta\,\xi\,dy\,(\cos^2\theta d\psi+\sin^2\theta d\phi)}{k\bar f\sqrt{\bar h_1\bar h_5}}+\frac{a^4\,s\,(1+s)\,\eta^2\,\xi^2(\cos^2\theta d\psi^2+\sin^2\theta d\phi^2)}{k^4\bar f\sqrt{\bar h_1\bar h_5}}\\
    &-\frac{2a\sqrt{Q_1Q_5}(r^2+b^2\xi\cos^2\theta)\,(dt-dy)\,(\gamma_1\cos^2\theta d\psi+\gamma_2\sin^2\theta d\phi)}{\bar f\sqrt{\bar h_1\bar h_5}(r^2+b^2\cos^2\theta)}\,,
    \end{aligned}
   \end{equation}
    \begin{equation}
    \begin{aligned} \nonumber
    C_2\,=\,&-Q_1\frac{dt\^ dy}{h_1 \bar f}+\frac{Q_5\cos^2\theta}{\bar h_1 \bar f}\Big(Q_1+r^2\xi+b^2(s+1)\xi+\frac{b^2 Q_1 q \sin^2\theta}{r^2+b^2\cos^2\theta}\Big)d\phi\^d\psi\\
    &+\frac{q \,a\,\sqrt{Q_1Q_5}(r^2+b^2)\cos^2\theta}{k\bar h_1\bar f\,(r^2+b^2\cos^2\theta)}dt\^dy+\frac{q\,a\,\sqrt{Q_1Q_5}\,r^2\sin^2\theta}{k\bar h_1\bar f(r^2+b^2\cos^2\theta)}dy\^d\phi\\
    &-\sin^2\theta\Big(a\,\sqrt{Q_1Q_5}-\frac{q\,a\,b^2\sqrt{Q_1Q_5}\cos^2\theta}{r^2+b^2 \cos^2\theta}\Big)\frac{(\gamma_1\,dt+\gamma_2\,dy)\^d\phi}{\bar h_1 \bar f}    \\
    &-\cos^2\theta\Big(a\sqrt{Q_1Q_5}+\frac{q\,a\,b^2\,\sqrt{Q_1Q_5}\sin^2\theta}{r^2+b^2 \cos^2\theta}\Big)\frac{(\gamma_2\,dt+\gamma_1\,dy)\^d\psi}{\bar h_1 \bar f}\\
    &+\frac{a\,b^2\,s\,(1+s)\,\xi}{k\sqrt{Q_1Q_5}\,\bar h_1\bar f}\Big(Q_1dt\^\big(\sin^2\theta d\phi+\cos^2\theta d\psi \big)+Q_5dy\^\big(\sin^2\theta d\phi+\cos^2\theta d\psi \big)\Big)\,,
    \end{aligned}
\end{equation}
\begin{equation}
\begin{aligned} \nonumber
    e^{2\Phi}&=\frac{\bar h_1}{\bar h_5}\,,
    \end{aligned}
\end{equation}
where 
\begin{equation}
    \bar f=\xi f\,,\qquad b^2=\frac{a^2\,\eta}{k^2}\,.
\end{equation}
We note that in the limit $s=0$, this solution reduces to that in Eq.~\eqref{superstube+sw}; in this limit $\bar f, \bar h_1$ and $\bar h_5$ reduce to $\bar f_{(0)},\bar h_{1(0)}$ and $\bar h_{5(0)}$. 

In our new solutions the regularity constraints~\eqref{GH constraint 3} are satisfied only by the coefficients of the pole at $r_+=0$ and not by the coefficients of the pole at $r_-=0$  in~\eqref{coefficients GLMT+sw}. This is as it should be, since the solution has a shockwave singularity at $ \bar f_{(0)}=0$, i.e.~at $(r=0,\theta=\pi/2)$.

The relation which ensures the absence of CTCs for smooth solutions, Eq.~\eqref{GH constraint 4}, is also not satisfied. Therefore we investigate the conditions for absence of CTCs directly. We do so by completing the squares in the periodic coordinates $(y,\phi,\psi)$ and by checking that the overall coefficient is globally non-negative. We first analyze the solution~\eqref{GLMT+sw Q1Q5} in the decoupling limit, where the form of the metric is simple enough to perform an analytic analysis.
Since the $g_{\phi\phi}$ and $g_{\psi\psi}$ are not affected by the spectral flow transformation~\eqref{GLMT SF}, we complete the squares in the following order: first $\phi$, then $\psi$, and finally $y$. In doing so, the conditions for absence of CTCs are independent of the spectral flow parameters $\gamma_1,\gamma_2$. We obtain the conditions
\begin{align}
\begin{aligned}
    &\phi \text{ coordinate:}\qquad\quad~ \sin^2\theta+\frac{q\sin^4\theta}{k^2r^2+\cos^2\theta}\;\geq\;0 \,, \\\
    &\psi \text{ coordinate:}
    \quad~~ k^2r^2\cos^2\theta+(1-q)\cos^4\theta\;\geq\;0 \,, \\
    &y \text{ coordinate:}\qquad\quad~~ \frac{(1-q)(k^2r^2+\cos^2\theta)}{k^2 r^2+(1-q)\cos^2\theta}\;\geq\; 0 \,,
\end{aligned}
\end{align}
which are always satisfied for $0\leq q < 1$. 

For the full asymptotically flat solution~\eqref{GLMT+sw Q1Q5}, as is often done we have performed a numerical analysis, based on which we can rule out CTCs with a high level of confidence.

Note that in our spectral flowed supertube solutions with shockwaves, Eq.~\eq{GLMT+sw Q1Q5}, the evanescent ergosurface is located at $f=0$, where $f$ is given in Eq.~\eqref{glmt-extras}. By contrast, the shockwave is located at $(r=0,\theta=\pi/2)$ which is not on the evanescent ergosurface for $s\neq 0$. Correspondingly, for $s\neq 0$ the addition of the shockwave does not come at zero cost in energy, and indeed we will now see that the momentum charge $Q_p$ is modified.

We now record the conserved quantities of our solutions \eqref{GLMT+sw Q1Q5}. As usual we wish to compare with five-dimensional D1-D5-P BPS black holes~\cite{Strominger:1996sh,Breckenridge:1996is}, so we are interested in the five-dimensional conserved mass and angular momenta obtained after dimensional reduction along the $y$ direction. These quantities are computed in Appendix \ref{charges11} and are given by Eqs.~\eqref{massADM} and~\eqref{ADMangmom}, which we record here as
\begin{align}
\begin{aligned}
M_{ADM}
&\,=\,\frac{\pi }{4 G_5} \left(Q_1+Q_5+\frac{Q_1 Q_5}{R_y^2}\frac{ s (s+\xi)}{k^2}\right)\,,\\
J^3\;=\;\frac{1}{2} (J^{\phi}-J^{\psi})
&\,=\,\frac{1}{2}\frac{ \xi N}{ k }+\frac{ s N}{ k }\,,\\
\bar J^3\;=\;\frac{1}{2} (J^{\phi}+J^{\psi})
&\,=\,\frac{1}{2}\frac{ \xi N}{ k } \,.
\label{ADM-glmt}
\end{aligned}
\end{align}

The condition $0\leq q < 1$ has a natural interpretation in the holographically dual CFT, as we shall see in the next section. Although the value $q=1$ is excluded, and the natural regime is small (but not infinitesimal) $q$, let us comment here on the form of the solutions as they approach the singular limit $q \to 1$ ($\xi \to 0$) with $\bar r=\sqrt{\xi} r$ fixed.
As $q \to 1$, our solutions approach small rotating D1-D5-P (BMPV~\cite{Breckenridge:1996is}) black holes, where here `small' means zero horizon size in supergravity. In the AdS$_3$ limit, the fractional spectral flow transformation \eqref{GLMT SF} relates these solutions to the AdS$_3$ limit of the two-charge D1-D5 BPS (non-rotating) small black hole solution. Similarly, this two-charge black hole solution is approached in the $q \to 1$ limit of the two-charge solutions with shockwaves \eqref{superstube+sw}. It is known that the two-charge black hole solution does not correspond to a microscopic profile function (or superposition of such functions), as discussed in~\cite{Lunin:2001jy,Mathur:2005zp,Mathur:2018tib}. These small black hole solutions are approached here because the $q \to 1$ limit is a singular limit which effectively coarse grains over all the microscopic details of the bound state; we shall elaborate on this in the next section once we have proposed the holographic description of these solutions.

\section{Holographic description of shockwave solutions}
\label{sec4}

In this section we identify a family of states of the D1-D5 orbifold CFT and propose that these are holographically dual to the AdS$_3\times$S$^3$ limits of the supergravity solutions~\eqref{superstube+sw} and~\eqref{GLMT+sw Q1Q5}. We perform tests of this proposal, including a precision holographic test, finding agreement.

\subsection{D1-D5 CFT}
\label{sec41}

We now briefly introduce the D1-D5 orbifold CFT. We consider D1-D5 bound states in Type IIB string theory, as described at the start of Section \ref{sec2}. Let the integer numbers of D1 and D5 branes be $n_1$ and $n_5$ respectively. In the AdS$_3\;\!\times\;\! $S$^3\times T^4$ decoupling limit, the holographically dual CFT is conjectured to be a two dimensional, $(4,4)$ SCFT with central charge $c=6 n_1n_5\equiv 6N$~\cite{Maldacena:1997re}. There is considerable evidence that there is a locus in moduli space where this theory becomes a symmetric product orbifold theory of $N$ copies of a $(4,4)$ free SCFT with target space $T^4$ and central charge $c=6$, see e.g.~\cite{Vafa:1995zh,deBoer:1998ip,Larsen:1999uk,David:2002wn} and references within.

We label the different copies of the symmetric product orbifold theory with the index $r=1,2,...,N$. The R-symmetry group is $SU(2)_L\times SU(2)_R$: we label indices in the respective fundamental representations by $\alpha,\dot \alpha=\pm$, and those in the adjoint with $a,\dot a=\pm,0$. It is also useful to label fields in terms of an organizational $SU(2)_C\times SU(2)_A\sim SO(4)$: it descends from the symmetry group of rotations in the four direction of the internal manifold, which is broken by the compactness of $T^4$. We use indices $A,\dot A=1,2$ for the fundamental of $SU(2)_C$ and  $SU(2)_A$ respectively. Each copy of the $c=6$ SCFT contains four free bosons $X_{(r)}^{A\dot A}$, four left-moving and four right-moving fermions $\psi_{(r)}^{\alpha,\dot A}$, $\bar \psi_{(r)}^{\dot \alpha,\dot A}$.

Being a symmetric product orbifold CFT, the theory contains twisted sectors. The twist operators are in one-to-one correspondence with the conjugacy classes of the permutation group $S_N$. 
These operators change the boundary conditions of the fields: for example, the boundary conditions corresponding to the permutation $(12...k)$ are given (on the cylinder) by
\begin{equation}\label{boundary condition twist}
\begin{aligned}
    X_{(1)}&\rightarrow X_{(2)}\rightarrow ...\rightarrow X_{(k)}\rightarrow X_{(1)}\,,\\
    \psi_{(1)}&\rightarrow \psi_{(2)}\rightarrow ...\rightarrow\psi_{(k)}\rightarrow \pm \psi_{(1)}\,,
\end{aligned}
\end{equation}
and analogously for the right-moving fermions. The $\pm$ boundary conditions in~\eqref{boundary condition twist} on the cylinder correspond respectively to the  R and NS sectors of the theory on a local covering space~\cite{Lunin:2001ne,Lunin:2001pw}; the lowest-dimension (`bare') twist operator corresponds to the NS-NS vacuum in the covering space. For more detailed discussion of this point, see~\cite{Chakrabarty:2015foa}. In the full symmetric product orbifold theory, twist operators are obtained by symmetrizing over all permutations in a given conjugacy class.

Given a state involving a collection of twist operators of cycle lengths $k_i$, it is common to describe the state as a collection of effective `strands' of lengths $k_i$. Strands of length $k_i$ can occur with multiplicity $N_i$, subject to the `strand budget' constraint $\sum_i N_i k_i=N$. 

As a first example, consider the state consisting of $N/k$ identical strands of length $k$, each in the lowest dimension state in the $k$-twisted sector. We denote this state by
\begin{equation}\label{twisted NS vacuum}
    \ket{0}_k^{N/k}=\ket{0}_k^{(1)}\otimes \ket{0}_k^{(2)}\otimes \cdots \otimes\ket{0}_k^{(N/k)}\,, 
\end{equation}
and we refer to it as the $k$-twisted NS vacuum. This state is an eigenstate of the left and right Virasoro modes $L_0,\bar L_0$ with eigenvalue $h=\frac{c}{24}(1-\frac{1}{k^2})$, it is a singlet under the $SU(2)_L\times SU(2)_R$ R-symmetry group and it is holographically dual to the $k$-orbifolded global AdS$_3\times$S$^3$ solution given in Eq.~\eq{twisted vacuum}.

Upon mapping twisted states into the local $k$-fold covering space~\cite{Lunin:2001ne,Lunin:2001pw}, there are no longer any twist operator insertions and the original $k$ copies of the fields in~\eqref{boundary condition twist} are mapped into single-valued fields.
In the $k$-fold covering space, the dimension $h_c$ and central charge $c_c$ are related to those in the physical CFT via $h=h_c/k$ and $c=k \:\! c_c$. Moreover, the $k$-twisted sector of the physical CFT contains fractional modes $n/k$ (and $(n+1/2)/k$), which correspond to integer modes $n$ (half-integer modes $n+1/2$) in the covering space. 

Our main interest is in black hole microstates in the RR sector of the theory, which arises directly from the AdS$_3$ decoupling limit of asymptotically flat configurations~(see e.g.~\cite{Aharony:1999ti}). 
One can map the NS sector of the CFT into the R sector using spectral flow~\cite{Schwimmer:1986mf}. 
Starting with a state of left scaling dimension $h$ and $SU(2)_L$ $J^3$ charge $m$ and acting with a left spectral flow transformation with parameter $\nu$, we obtain a state in the same twist sector with left dimension and charge $(h',m')$ given by
\begin{equation}\label{SF h j}
    h'=h+2\nu m+\frac{c \nu^2}{6}\,,\qquad m'=m+\frac{c\nu}{6}\,.
\end{equation}
When $\nu$ is half integer, a spectral flow transformation maps a state in the NS sector to a state in the R sector. 
When considering spectral flow  of the full CFT, we have $c=6N$. If we consider an individual strand of length $k$, we have $c=6k$. A similar transformation holds for the right sector of the theory, with parameter $\bar \nu$.

When $(\nu,\bar \nu)=(\frac{1}{2},\frac{1}{2})$, the untwisted NS vacuum $\ket{0}_1^N$ is mapped into a RR state with $h=\bar{h}=N/4$, which is therefore a RR ground state. It carries R-symmetry charge $m=\bar m=N/2$ and we shall denote it with $\ket{++}_1^N$. The other RR ground states can be obtained from spectral flow of other anti-chiral primaries (i.e. operators satisfying the bound $h = j = -m$, $\;\bar h = \bar j = -\bar m$) by applying the same spectral flow transformation. For a given twist $k$ there are (anti-)chiral primaries of dimension $h=k/2$, $\,h=(k-1)/2$ and $h=(k+1)/2$.

Let us now consider the sector of the full CFT composed of $N/k$ strands of length $k$. In this sector, there is an enhancement of spectral flow known as \textit{fractional} spectral flow~\cite{Martinec:2001cf,Martinec:2002xq},~\cite{Giusto:2012yz,Chakrabarty:2015foa}. This operation is naturally thought of as ordinary spectral flow in the $k$-fold covering space and means that the values $\nu \in \mathbb{Z}/k$ give rise to physical states in the same (R or NS) sector of the theory, while the values $\nu \in (\mathbb{Z}+\tfrac{1}{2})/k$ map from R to NS in the $k$-fold cover.

The backgrounds to which we add shockwaves in this work are the heavy BPS RR states obtained by chiral fractional spectral flow of the state $\ket{++}_k^{N/k}$, studied in~\cite{Giusto:2012yz}. Specifically, we consider $\ket{++}_k^{N/k}$ as our reference state and perform left fractional spectral flow with parameter $\nu=s/k$. These states were proposed to be holographically dual to the bulk configurations in Eqs.~\eqref{GLMT}--\eqref{glmt-extras} in~\cite{Giusto:2012yz} and this proposal has passed non-trivial holographic tests~\cite{Giusto:2012yz,Galliani_2016}.
We shall exhibit these CFT states in more detail in Section \ref{sec:CFT GLMT}.

\subsection{Holographic description of shockwaves in supertube backgrounds}\label{sec:CFT supertube+sw}

The first shockwave solution we reviewed, in Eq.\;\eqref{sw+twisted vacuum}, for $k=1$ describes a shockwave on the global AdS$_3\times$S$^3$ vacuum. As we have discussed, the shockwave describes the backreaction of a distribution of high-energy massless particles. Supergravity excitations on the vacuum are holographically dual to CFT states in short multiplets whose top (bottom) component is a chiral (anti-chiral) primary, see e.g.~\cite{Aharony:1999ti,Rawash:2021pik}. In our conventions, the shockwave of \eqref{sw+twisted vacuum} is holographically dual to a set of several anti-chiral primaries of the dual CFT with large conformal dimension and R-charge, and therefore high twist~\cite{Lunin:2002bj}.

Upon spectral flow to the RR sector, (anti-)chiral primaries transform into RR ground states. 
Suitably coherent RR ground states of the D1-D5 system can be described in terms of eight profile functions $g_i(v')$ in $\mathbb{R}^8$, where $v'$ is a null coordinate, with periodicity $L=2\pi Q_5/R_y$~\cite{Lunin:2001jy,Lunin:2002iz,Kanitscheider:2007wq}.

Let us consider the twisted circular supertube geometry that is generated by a circular profile of radius $a/k$ in the $x_1$-$x_2$ plane,
\begin{equation}\label{profile function supertube}
    g_1(v')+i g_2(v')=\frac{a}{k} \,e^{\frac{2\pi i k}{L}v'}\,,\qquad g_{i\neq 1,2}=0\,.
\end{equation}
The dictionary between the profile and the CFT state can be found in~\cite{Lunin:2001jy,Lunin:2002iz,Kanitscheider:2007wq,Giusto:2015dfa} (see also~\cite{Giusto:2019qig} for clarification of some details). The CFT state dual to the microstate generated by the profile~\eqref{profile function supertube} is
\begin{equation}\label{CFT twisted supertube}
    \ket{++}_k^{N/k}\,.
\end{equation}

Let us now consider the AdS$_3\times$S$^3$ limit of the solution with shockwave in Eq.~\eqref{superstube+sw}. 
If we switch off the shockwave by setting $q=0$, this solution is the one corresponding to the profile~\eqref{profile function supertube} and CFT state~\eqref{CFT twisted supertube}. For non-zero $q$, this solution can be generated by an approximate profile function by performing two steps (see \cite[Fig.~2]{Lunin:2002bj}
for a pictorial representation).  
The first step is to consider a profile which initially traverses, $k$ times, a circle of radius $\bar a/k=\xi a/k$ in the $x_1$-$x_2$ plane on the interval $v'\in [0,\xi L]$, and which then remains in the same $x$-location for the remainder of its length (recall $\xi=1-q$):
\begin{equation}\label{profile supertube+sw before smearing}
\begin{aligned}
    g_1(v')+i g_2(v')&\,=\,\frac{\bar a}{k} e^{\frac{2\pi i k}{\xi L}v'}\,,\qquad  0\leq v'\leq \xi L\\
    g_1(v')+i g_2(v')&\,=\,\frac{\bar a}{k} \,,\hspace{4.2em}  \xi L\leq v'\leq  L\\
    g_{i\neq 1,2}&\,=\,0\,.
\end{aligned}
\end{equation}
The constant segment represents the high-twist chiral primaries, corresponding to profile Fourier modes with high mode numbers and small amplitudes that are not resolved by supergravity.

The second step is to break this constant segment into several smaller segments and smear over their locations within the overall profile to obtain a uniform distribution (subject to additional conditions described in detail in~\cite{Lunin:2002bj}). The resulting approximate profile reproduces the supergravity solution with shockwave given in Eq.~\eqref{superstube+sw}~\cite{Lunin:2002bj}. This procedure is most natural when $q$ is small compared to 1 (but not infinitesimally small).

We now discuss the holographic description of these solutions, refining the discussion in~\cite{Lunin:2002bj} given for $k=1$.
The circular segment of the profile function \eqref{profile supertube+sw before smearing} corresponds to a set of strands of type $\ket{++}_k$. The constant segment that is smeared corresponds to some collection of RR ground state strands whose strand lengths are large in a sense that we will make precise shortly. The polarizations of the RR strands are not resolved in supergravity; for concreteness we will take them to be the five bosonic RR ground states that are invariant on the $T^4$, commonly labelled by their R-charges as $\ket{\varepsilon\bar\varepsilon}=\ket{\pm\pm},\ket{\pm\mp},\ket{00}$.
As a first pass, we write this family of CFT states as follows (and arbitrary superpositions thereof):
\begin{equation}\label{superstube+sw CFT pre}
    \ket{++}_k^{N_0}
    \ket{\varepsilon_1\bar\varepsilon_1}_{k_1}^{d_1}
    \cdots
    \ket{\varepsilon_{n_s}\bar\varepsilon_{n_s}}_{k_{n_s}}^{d_{n_s}}\,,\qquad \frac{k_i}{k} \in \mathbb{Z}, \qquad \frac{k_i}{k} \gg 1\,,\qquad
    N_0 k+\sum_{i=1}^{n_s} d_i k_i=N\,.
\end{equation}
Here $N_0$ is the number of strands representing the supertube background, and $d_i$ is the degeneracy of the various strands making up the shockwave. We work at leading order in large $N$. We take the parameter $k$ to be independent of $N$, so that $N_0\sim N$. We also take $q$ and $\xi$ to be independent of $N$. For ease of terminology we shall refer to the strands of length $k_i$ as the long strands, and to those of length $k$ as the short strands.

In the long strand sector, neither the parameters $k_i$, $d_i$, $n_s$, nor the distribution of polarizations are fixed. This is the CFT analog of the fact that in the bulk the total energy of the shockwave is known, however it is not known how this energy is distributed among the high-energy supergravity quanta making up the shockwave.

Each segment of the supergravity profile \eq{profile supertube+sw before smearing} corresponds to a component of the dual CFT state that contributes a finite fraction of the total strand budget at large $N$. Considering the overall strand budget of the set of all long strands, we must also have 
$\sum_i d_i k_i \sim N$.

We will shortly refine the above to derive that at leading order in large $N$ we must have $k N_0 = \xi N$ and thus $\sum_i d_i k_i  =  q N$. Thus $\xi$ will be the fraction of the total strand budget taken up by the short strands, and $q$ will be the fraction of the total strand budget taken up by the long strands.

The supergravity profile does not explicitly include any Fourier modes higher than $k$ with finite amplitude. From the two-charge dictionary as made precise in~\cite{Giusto:2015dfa}, 
this means that the CFT state cannot contain any long strands with both $k_i\sim N^0$ and $d_i\sim N$.
Therefore no $d_i$ can scale as $N$.
We shall derive a stronger condition shortly.

We now refine
the condition 
$k_i \gg k$ stated in~\cite{Lunin:2002bj} (for $k=1$). Our main analysis will involve a precision holography calculation. However it is instructive to make a brief crude first pass by temporarily making the simplifying assumption that the length of all the long strands scales in the same way, which we write as $k_i \sim N^b$, where a priori $0\le b \le 1$. Similarly we temporarily assume that all the degeneracies of the long strands scale as $d_i \sim N^d$ with $0\le d < 1$, recalling that we have excluded $d=1$ in the previous paragraph. Then the condition $\sum_i d_i k_i \sim N$ requires that $n_s \sim N^A$ with $b+d+A=1$ and a priori $0\le A \le 1$.

Now, in order for there to be enough different integers $k_i$ to have order $N^A$ types of long strands, we must have $b \ge A$. Combining this with the constraint $b+d+A=1$, we find
\begin{equation}
    A \;\le\; \frac{1-d}{2} \,, 
    \qquad 
    b \;\ge\; \frac{1-d}{2} 
    \quad~~ \Rightarrow \quad~~
    b > 0 \;.
\end{equation}
So in this simplified analysis, we see that the length of the long strands must scale with a positive power of $N$. Furthermore,
\begin{align}
\label{natural sw regime-00}
     \sum_{i=1}^{n_s} d_i 
      \;\sim\; N^{1-b} \qquad\text{with} \quad
    b > 0 \,.
\end{align}
This relation will be important for matching the conserved charges. Using precision holography we will shortly establish it in general, with no assumption on the scaling of the different $k_i$.

As a side comment, let us note that when we allow the different $k_i$ to scale as different powers of $N$, it is possible for some strand lengths to scale as $k_i\sim N^0$ with degeneracies that scale as $d_i\sim N^d$ with $d<1$, provided that $k_i \gg k$. Since such strands individually account for a vanishingly small strand budget at large $N$ (of order $N^d$), one would discard them unless the same is true for all the other long strands present, for instance if all $d_i k_i \sim N^d$ and $n_s\sim N^{1-d}$.
However in such a CFT state, the vast majority of types of strands will have lengths that scale as some positive power of $N$ (at least $N^{1-d}$).

\subsection{Precision holography analysis}

We now proceed to our precision holography analysis, in which we will prove for general $k_i$ that the condition \eqref{natural sw regime-00} is necessary. This condition will also be sufficient to ensure agreement between gravity and CFT to the precision we probe.

We use the holographic dictionary developed in~\cite{Kanitscheider:2006zf,Giusto:2015dfa,Giusto:2019qig,Rawash:2021pik}. 
Consider a heavy 1/4 or 1/8-BPS CFT state dual to a given bulk configuration, and a light operator $\mathcal{O}$ which is either a chiral primary or a descendant of a chiral primary under the global part of the chiral algebra.
Then the dictionary relates the expectation value of $\mathcal{O}$ to the asymptotic expansion of the supergravity field dual to $\mathcal{O}$.

We shall focus on a particular sector of the holographic dictionary. To keep the presentation concise, we shall describe the computation in outline, without a lengthy review. We record some definitions of chiral primary operators in Appendix~\ref{operators D1D5 CFT}, and for further details we refer the reader to~\cite{Rawash:2021pik}.

On the bulk side, we work in the AdS$_3$ decoupling limit. We expand fluctuations in S$
^3$ harmonics and consider a single-particle excitation that is a scalar in AdS$_3$. 
Since we are considering a two-charge configuration, the four-dimensional base space of the supergravity ansatz \eqref{ansatzSummary} is flat $\mathbb{R}^4$. We work in spherical polar coordinates in which it takes the form 
\begin{equation}\label{base metric}
    ds_4^2=d\bar r^2+\bar r^2(d\theta^2+\sin^2\theta d\phi^2+\cos^2\theta d\psi^2)\,,
\end{equation}
where we have labeled the radial coordinate by $\bar{r}$, for consistency with the notation used in the two-charge solution with shockwave in  Eq.~\eqref{superstube+sw}.

In these coordinates it is useful to  expand the harmonic functions $Z_1,Z_2$ that appear in the BPS ansatz in Appendix~\ref{app:sugra} in scalar S$^3$ harmonics $Y_\k^{m_\k,\bar m_\k}$ and for large $\bar{r}$ as follows:
\begin{equation}\label{eq:geometryexpansion} 
\begin{aligned}
 Z_1 &\,=\,\frac{Q_1}{\bar r^2} \bigg(1+ \sum_{\k=1}^2\; \sum_{m_\k,\bar{m}_\k=-\k/2}^{\k/2} a_0^{\k}\, f^{\,(m_\k,\bar m_\k)}_{1\k} \,\frac{Y_\k^{m_\k,\bar m_\k}}{\bar r^\k}+O( r^{-3})\bigg)\,,\\
 Z_2&\,=\,\frac{Q_5}{\bar r^2} \bigg(1+ \sum_{\k=1}^2\; \sum_{m_\k,\bar{m}_\k=-\k/2}^{\k/2} a_0^{\k}\, f^{(m_\k,\bar{m}_\k)}_{5\k} \,\frac{Y_\k^{m_\k,\bar m_\k}}{\bar r^\k}+O( r^{-3})\bigg)\,,
\end{aligned}
\end{equation}
 where $a_0=\frac{\sqrt{Q_1Q_5}}{R_y}$.

The particular AdS$_3$ scalar we consider is denoted $s^{(6)(a,\dot a)}_{\k=2}$. We denote the coefficient of $\bar r^{-2}$ in its large $\bar r$ expansion by $\Big[s^{(6)(a,\dot a)}_{\k=2}\Big]$, following the notation of~\cite{Rawash:2021pik}.  Choosing the gauge $f_{11}^{(m_1,\bar m_1)}+f^{(m_1,\bar m_1)}_{51}=0$, one then  has~\cite{Kanitscheider:2006zf,Rawash:2021pik}
\begin{equation}\label{s6k2}
     \Big[s^{(6)(a,\dot a)}_{\k=2}\Big]\,=\,\sqrt{\frac{3}{2}} \Big(f_{12}^{(a,\dot a)}-f_{52}^{(a,\dot a)}\Big)\,.
\end{equation}
The explicit values of the harmonic functions characterizing the backreaction of shockwave on a supertube background were obtained in
\cite[Eq.\;(3.18)]{Lunin:2002bj}. Changing coordinates to recast the base metric into the form~\eqref{base metric}, performing the asymptotic expansion in~\eqref{eq:geometryexpansion} and using \eqref{s6k2}, one obtains that the AdS$_3$ limit of the solution describing a two-charge supertube with shockwave in Eq.~\eqref{superstube+sw} has the property that
\begin{equation}\label{s6k2 sw}
    \Big[s^{(6)(a,\dot a)}_{\k=2}\Big]\,=\,0\,.
\end{equation}

On the CFT side, the dual operator is a scalar chiral primary operator of dimension two and we shall denote it by $\tilde{\Sigma}_3^{a\dot a}$, again following the notation of~\cite{Rawash:2021pik}. This operator is composed of a linear combination of single-trace operators of dimension two and double-trace operators made up of dimension one operators.
Truncating this operator appropriately to the supergravity ansatz in which we work, its explicit form is:
 \begin{equation}\label{tilde Sigma_3}
     \tilde\Sigma_3^{a \dot a}\,\equiv\, \frac{3}{2} \left[\left(\frac{\Sigma_3^{a \dot a}}{N^{\frac32}}-\frac{\Omega^{a \dot a}}{3N^{\frac12}}\right)+\frac{1}{N^{\frac12}}\left(-\frac{2}{3}(\Sigma_2\cdot \Sigma_2)^{a \dot a}
     +\frac{1}{3}(J\cdot \bar{J})^{a \dot a}\right)\right]\,,
 \end{equation}
 where the operators entering this linear combination are defined in Appendix~\ref{operators D1D5 CFT}.
In this sector, the dictionary reads~\cite{Giusto:2019qig,Rawash:2021pik}
 \begin{equation}\label{holographic dictioanry Sigma3}
 \big\langle\tilde{\Sigma}_3^{a\dot a}\big\rangle\;=\;(-1)^{a+\dot a}\frac{\sqrt{N}}{\sqrt{2}}\Big[s^{(6)(-a,-\dot a)}_{\k=2}\Big]\,.
 \end{equation}

Combined with the result in Eq.~\eqref{s6k2 sw}, this implies that the dual CFT state~\eqref{superstube+sw CFT pre} must have a vanishing expectation value of the operator $\tilde\Sigma_3^{a \dot{a}}$. This requirement will yield the claimed constraint~\eqref{natural sw regime-00}.

For ease of presentation, we shall make two simplifications: first, we take the twist parameter in \eqref{superstube+sw CFT pre} to be $k=1$ for the remainder of this subsection, and second, we focus on CFT states involving only strands of polarization type $\ket{++}$. The computation and result for generic $k$ and generic long strand polarizations are entirely analogous. A more general case involving both $\ket{++}$ and $\ket{--}$ polarizations for the long strands is described in Appendix~\ref{app: precision holography}.

We shall focus on a particular component, specifically the operator $\tilde{\Sigma}_3^{00}$. Among the operators that mix in Eq.~\eqref{tilde Sigma_3}, there are three operators that have a non vanishing expectation value on the class of states~\eqref{superstube+sw CFT pre}: the single-trace operators $\Sigma_3^{00}$, $\Omega^{00}$ and the double-trace $\big(J\cdot \bar J\big)^{00}$. The contribution of the other double-trace operator is subleading in $N$, so we shall ignore it.

First, we analyze the contribution from the twist-three operator $\Sigma_3^{00}$. This operator acquires a non-vanishing expectation value by mapping two strands of different length into themselves, permuting the copies~\cite{Giusto:2019qig}. The fusion coefficient of the process can be computed holographically; we describe the computation in Appendix~\ref{app:fusion sigma3}. The result is:
\begin{equation}\label{Sigma3 k1k2}
    \sigma_3^{00}\ket{++}_{k_1}\ket{++}_{k_2}\,=\,\frac{(k_1+k_2)^2}{6 k_1^2 k_2^2}\big(1-\delta_{k_1,k_2}\big)\ket{++}_{k_1}\ket{++}_{k_2}\,.
\end{equation}
The expectation value of $\Sigma_3^{00}$ on the full state~\eqref{superstube+sw CFT pre} arises from the process
\begin{equation}\label{process Sigma_3}
    \Sigma_3^{00}\Big(\ket{++}_1^{N_0}\prod_i\ket{++}_{k_i}^{d_i}\Big)\,=\,\bigg(\sum_{i\neq j}\frac{(k_i+k_j)^2}{6 k_i k_j} d_i d_j+\sum_i\frac{(k_i+1)^2}{6 k_i} N_0 d_i\bigg)\Big(\ket{++}_1^{N_0}\prod_i\ket{++}_{k_i}^{d_i
    }\Big)\,.
\end{equation}
The two terms in the first parenthesis after the equality sign correspond respectively to the processes in which the twist-three operator acts on two long strands, and on a long and a short strand. Let us consider the first contribution: it is given by combining~\eqref{Sigma3 k1k2} with the fact that $\Sigma_3$ can act on any of the $d_i d_j$ pairs of strand of different length and can cut each of them in $k_i$ and $k_j$ different positions. The second contribution works analogously.

Second, we analyze the operator $\Omega^{00}$.
The states $\ket{++}_k$ are eigenstates of this operator with the following eigenvalue~\cite[Eq.\;(5.40)]{Giusto:2019qig}
\begin{equation}\label{eigenvalue Omega00}
    \Omega^{00}\ket{++}_k\;=\;\frac{1}{2k}\ket{++}_k\,.
\end{equation}
Therefore the operator  $\Omega^{00}$ acquires a non-vanishing expectation value via the process
\begin{equation}\label{process Omega}
    \Omega^{00}\Big(\ket{++}_1^{N_0}\prod_i\ket{++}_{k_i}^{d_i}\Big)\,=\,\bigg(\frac{N_0}{2}+\sum_i\frac{d_i}{2 k_i} \bigg)\Big(\ket{++}_1^{N_0}\prod_i\ket{++}_{k_i}^{d_i}\Big)\,.
\end{equation}

Third, we consider the double-trace operator $\big(J\cdot \bar J\big)^{00}$. Its expectation value arises from the process
\begin{equation}\label{process JJ}
   \big(J\cdot \bar J\big)^{00}\Big(\ket{++}_1^{N_0}\prod_i\ket{++}_{k_i}^{d_i}\Big)\,=\,
   \frac{2}{N}\bigg(\frac{N_0^2}{4}+2 \frac{N_0}{2}\sum_i\frac{d_i}{2}+\sum_{i,j}\frac{d_i d_j}{4} \bigg)\Big(\ket{++}_1^{N_0}\prod_i\ket{++}_{k_i}^{d_i}\Big)\,.
\end{equation}
The three terms after the equality sign correspond respectively to: (i) the action of both the left and the right current on a short strand; (ii) one current acting on a short and one on a long strand; and  (iii) both currents acting on a long strand.
By combining Eqs.~\eqref{process Sigma_3}--\eqref{process JJ} we obtain the expectation value of the single-particle operator:
\begin{equation}\label{VeV tilde Sigma3}
\begin{aligned}
    \big\langle\tilde\Sigma_3^{00}\big\rangle&\,=\,\frac{3}{2N^{3/2}}\bigg[\frac{2}{3}N_0\sum_i d_i+\sum_{i\neq j}d_id_j\frac{(k_i+k_j)^2}{6k_i k_j}-\frac{1}{6}\sum_{i,j}d_id_j\Big(\frac{k_i}{k_j}-1\Big)
    \bigg]\\
    &\,=\,\frac{1}{N^{3/2}}\bigg[N_0\sum_i d_i+\sum_{i\neq j}d_id_j\frac{k_j^2+3k_ik_j}{4k_i k_j}\bigg]\,,
    \end{aligned}
\end{equation}
where the last equality follows by noticing that the $i=j$ parts of the last term of the first line vanish. 

We are using the normalization of the holographic dictionary employed in~\cite{Giusto:2019qig,Rawash:2021pik}, in which the contribution of an operator is visible in the supergravity approximation if its expectation value is of order $N^{1/2}$ in the large $N$ limit. Therefore the expectation value of $\tilde\Sigma_3^{00}$ will agree with Eq~\eqref{s6k2 sw} if and only if its large $N$ scaling is subleading with respect to $ N^{1/2}$. 

We note that Eq.~\eqref{VeV tilde Sigma3} is the sum of two positive terms, so no cancellation can occur. Let us thus consider the first term. We have $N_0 \sim N$ and therefore we require that
\begin{align}
\label{natural sw regime-0}
     \sum_{i=1}^{n_s} d_i \;\sim\; N^{1-\alpha} \qquad\text{for~some} \quad
    \alpha > 0 \,.
\end{align}
We emphasize that we have now established that this condition is necessary in general, for any set of long strand lengths $k_i$.

Next we consider the second term. Again as a crude first pass, suppose that all the various $k_i$ scale as the same power of $N$. Then an upper bound on the scaling of this term is $N^{2(1-\alpha)}$ with $\alpha > 0$, from squaring~\eq{natural sw regime-0}. Then this term, and thus the total expectation value, are subleading compared to $N^{1/2}$ as required.

More generally, suppose instead that there are different values of $k_i$ scaling as different powers of $N$. The term corresponding to $3k_ik_j$ in the numerator of the second line of \eqref{VeV tilde Sigma3} is subleading
compared to $N^{1/2}$
by the same argument as in the last paragraph. An upper bound on the remaining term is given by adding in the $i=j$ terms into the sum, obtaining
\begin{equation} \label{eq:bound}
    \frac{1}{N^{3/2}}\bigg(\sum_j d_j k_j\bigg) \bigg(\sum_i\frac{d_i}{k_i}\bigg)\,.
\end{equation}
The first sum is of order $N$, while the second is bounded above by $\sum_{i} d_i \sim N^{1-\alpha}$. So~\eqref{eq:bound} is also subleading compared to $N^{1/2}$. Therefore the condition~\eqref{natural sw regime-0} is also sufficient to ensure that the precision holographic test is passed.

We now use the condition \eq{natural sw regime-0} to determine $N_0$, the degeneracy of the twist-$k$ strands, in the large $N$ limit. The analysis of the conserved charges of the metric~\eqref{superstube+sw} in~\cite{Marolf:2016nwu} established that the angular momentum carried by the solution describing a shockwave on a supertube background is suppressed by a factor of $\xi$ with respect to that of the supertube solution:
\begin{equation}
    \big\langle J^3\big\rangle_{\text{Supertube+SW}}\,=\,\xi \big\langle J^3\big\rangle_{\text{Supertube}}\,=\,\xi \frac{N}{2k}\,.
\end{equation}
The same value is obtained upon setting $s=0$ in the conserved charges in Eq.~\eqref{ADM-glmt}.
The CFT state~\eqref{superstube+sw CFT pre} is an eigenstate of the current operator $J^3$, with eigenvalue:
\begin{equation}
    \big\langle J^3\big\rangle\,=\,\frac{N_0}{2}+\sum_{i=1}^{n_s}\varepsilon_i \frac{d_i}{2}\,.
\end{equation}

Recall that we have taken $k \sim N^0$ and $N_0\sim N$. We have just shown that $\sum d_i \sim N^{1-\alpha}$ with $\alpha>0$. So at large $N$ the contribution of the long strands to the expectation value of $J^3$ is subleading. As anticipated above, we thus conclude that at leading order in large $N$,
\begin{equation}\label{N0 N relation}
    N_0\,=\,\xi \frac{N}{k}\,.
\end{equation}
Therefore, as claimed, $q$ is the fraction of the total strand budget taken up by the long strands, and $\xi=1-q$ is the fraction of the strand budget taken up by the short strands.

For convenient reference we now record the more refined version of the family of CFT states in Eq.~\eqref{superstube+sw CFT pre} as
\begin{align}
\begin{aligned}
\label{superstube+sw CFT final}
    &\ket{++}_k^{N_0}
    \ket{\varepsilon_1\bar\varepsilon_1}_{k_1}^{d_1}
    \cdots
    \ket{\varepsilon_{n_s}\bar\varepsilon_{n_s}}_{k_{n_s}}^{d_{n_s}}\,,\qquad~ \frac{k_i}{k} \in \mathbb{Z},
    \qquad~ \frac{k_i}{k} \gg 1\,,
    \cr &\qquad\qquad\qquad\quad~~ k N_0\,=\,\xi N \,, \qquad \sum_{i=1}^{n_s} d_i k_i\,=\, qN\,, \qquad 
      \sum_{i=1}^{n_s} d_i \,\sim\, N^{1-\alpha} \,, \qquad 
    \alpha > 0 \,.
    \end{aligned}
    \end{align}
We remind the reader that while the presence of the shockwave decreases the angular momentum, the total energy of the system is left unchanged and is given by $h=\bar h=\frac{N}{4}$.

Let us return to the condition $0\leq q < 1$ derived in Section \ref{sec3}. We make two brief observations here that shed further light on the condition $q < 1$. First,  the string profile~\eqref{profile supertube+sw before smearing} would become a straight line in the limit $q \to 1$, which is microscopically inconsistent with the fact that the configuration carries two charges (see e.g.~\cite{Mathur:2005zp}). Second, the family of CFT states \eqref{superstube+sw CFT pre} involves long strands of winding $k_i \gg k$ whose details are not resolved by supergravity relative to the short strands of length $k$. In the limit $q \to 1$, the short strands are no longer present, so the approximation of a smeared profile is no longer valid. For such CFT states a more refined bulk description is required, and is given by the extrapolation of the general two-charge microstate solutions into the stringy regime~\cite{Lunin:2001jy,Lunin:2002bj,Taylor:2005db,Kanitscheider:2007wq}.

As a final comment on these microstates, we note that the proposed holographic description of the $k=1$ supertube background with shockwave is similar to the proposed holographic description of small two-charge BPS black rings of the D1-D5 system~\cite{Elvang:2004ds,Bena:2004tk,Balasubramanian_2005}, where again here `small' means zero horizon area in supergravity. It would be interesting to further investigate this similarity.

\subsection{Holography of fractionally spectral flowed supertubes}\label{sec:CFT GLMT}
In this section we review in more detail the holographic description of the fractionally spectral flowed supertube solutions~\cite{Giusto:2012yz} and discuss some of their physical properties. 

As mentioned at the end of Section~\ref{sec41}, the dual CFT states to the fractionally spectral flowed supertube solutions given in Eq.~\eqref{GLMT} are 1/8-BPS microstates obtained by left fractional spectral flow of the 1/4-BPS state $\ket{++}_k^{N/k}$ by an amount $\nu=s/k$ with $s\in \mathbb{Z}$. The spectral flow adds left-moving fermionic excitations, while leaving the right movers in the ground state; this results in a non-zero momentum charge $n_p=h-\bar h$. 
The state of each strand takes the explicit form
\begin{equation}\label{GLMT CFT}
    \ket{++}_{k,s}\equiv
    \begin{cases}
      \Big[\psi^{+1}_{-\frac{s}{k}}\psi^{+2}_{-\frac{s}{k}}\cdots \psi^{+1}_{-\frac{1}{k}}\psi^{+2}_{-\frac{1}{k}}\Big]\ket{++}_k\,,\qquad s\ge 1
     \\
     \\
      \Big[\psi^{-1}_{\frac{s+1}{k}}\psi^{-2}_{\frac{s+1}{k}}\cdots \psi^{-1}_{0}\psi^{-2}_{0}\Big]\ket{++}_k=\Big[\psi^{-1}_{\frac{s+1}{k}}\psi^{-2}_{\frac{s+1}{k}}\cdots \psi^{-1}_{-\frac{1}{k}}\psi^{-2}_{-\frac{1}{k}}\Big]\ket{-+}_k\,,\qquad s \le -1 \,.
    \end{cases}
\end{equation}
Recall that in the $k$-twisted sector the level spacing of the excitations is in units of $1/k$. This means that spectral flow is the energetically most convenient way to add charge, corresponding to filling a Fermi sea of excitations up to the fractional level $s/k$ for $s\ge 1$, or the level $-(s+1)/k$ for $s\le 1$.
Fractional spectral flow has an entirely analogous effect on the other RR ground states with polarizations $\ket{--},\ket{\pm\mp},\ket{00}$; for further details see e.g.~\cite{Bena:2016agb}.

Let us record the charges of the state~\eqref{GLMT CFT}. The spectral flow transformation involves only the left sector of the theory, so the right charges are the same as those of the two-charge circular supertube. The left charges follow from Eq.~\eqref{SF h j} and are
\begin{equation}\label{CFT charges GLMT}
\begin{aligned}
    h&\,=\,\frac{N}{4}+\frac{Ns(s+1)}{k^2}\,, \qquad \bar h \,=\,\frac{N}{4}\,, \\ 
    m&\,=\,\frac{N}{k}\Big(s+\frac{1}{2}\Big)\,,\qquad\qquad \bar m\,=\,\frac{N}{2k}\,.
    \end{aligned}
\end{equation}
Importantly, not all values of $s, k$ are allowed. The momentum per strand $p$ is required to be an integer: 
\begin{equation}\label{eq:s-k-cond-GLMT}
p \,=\, \frac{s(s+1)}{k}\,\in\, \mathbb{Z}.
\end{equation}

\begin{figure}[t]
    \centering
    \includegraphics[width=15.1cm]{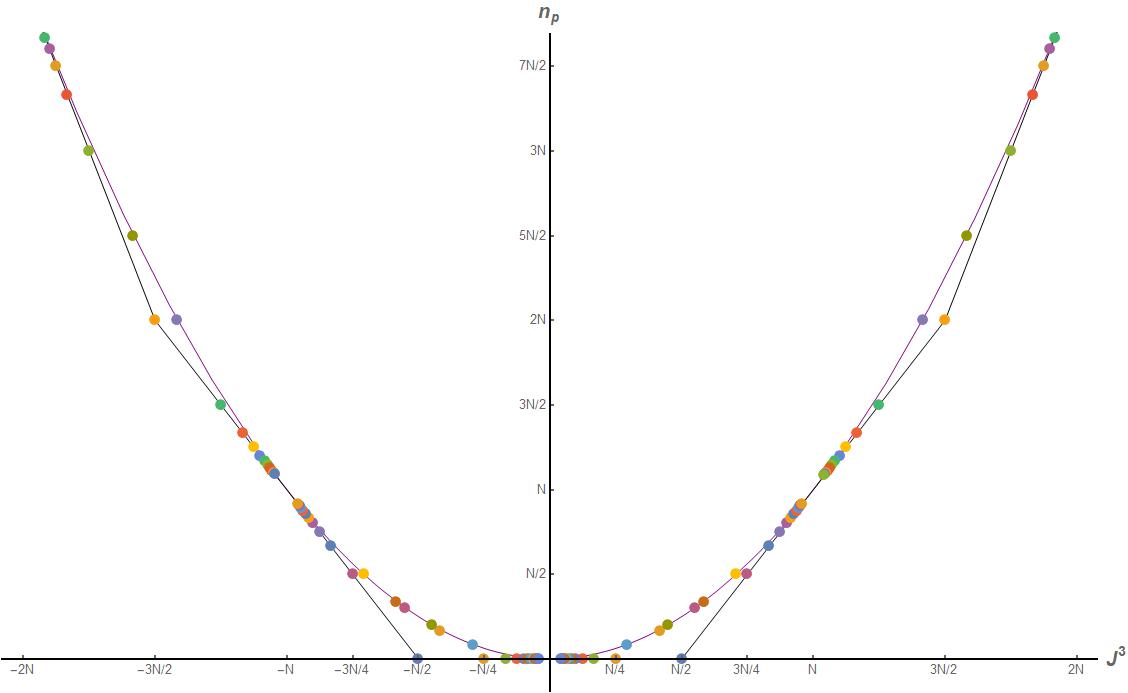}
    \caption{Quantum numbers $(J^3,n_p)$ for  fractional spectral flowed supertube states with $k \le 12$ and $|s|\le 12$ satisfying the condition \eqref{eq:s-k-cond-GLMT} and for which $|J^3|\le 2N$. All points lie outside the parabola, even though some appear very close to it.}
    \label{fig:GLMT}
\end{figure}

In Figure~\ref{fig:GLMT} we display the $(J^3,n_p)$ phase diagram for the D1-D5-P system in the RR sector. The black polygon represents the unitarity bound: allowed CFT states exist only on and above this threshold. The parabola $n_p=(J^3)^2/N$ delimits the region of existence of finite-size BMPV black holes, which exist only inside the parabola. Note that inside but very close to the parabola, the small BMPV black holes are sub-dominant to either a BMPV plus supertube or black ring~\cite{Bena:2011zw}. The fractionally spectral flowed supertube solutions live in the region bounded by the black polygon and the purple parabola. We represent with dots the solutions with $k \le 12$ and $|s|\le 12$. 

Note that the dots in the corners of the unitarity bound polygon are the states with $k=1$. In our conventions the interval $0<J^3<N/2$ with $n_p=0$ contains the RR ground states with $k>1$ and $s=0$, i.e.~the states $\ket{++}_{k}^{N/k}$. 
Dots in the interval $-N/2<J^3<0$ with $n_p=0$ correspond to $k>1$ and $s=-1$, which are the two-charge states $\ket{++}_{k,s=-1}^{N/k}=\ket{-+}_k^{N/k}$. Dots on the remaining lines of the polygon correspond to spectral flowed states that have $s/k \in \mathbb{Z}$ or $(s+1)/k \in \mathbb{Z}$. 

The remainder of the states are the most interesting physically. These lie closer to the BMPV parabola, and  have $k>1$ and neither $s/k \in \mathbb{Z}$ nor $(s+1)/k \in \mathbb{Z}$.
These were the states of primary interest in~\cite{Giusto:2012yz}.

\subsection{Holography of shockwaves in fractionally spectral flowed supertubes}\label{sec:CFT GLMT+sw}

We now propose the holographic description of the AdS$_3\times$S$^3$ limit of the solutions describing fractional spectral flowed supertubes with shockwaves in Eq.~\eqref{GLMT+sw Q1Q5}. 
The AdS$_3\times$S$^3$ limit of the metrics are given in Eq.~\eqref{eq:GLMT+sw-AdS}.

Recall that the spectral flow large coordinate transformation~\eqref{GLMT SF} maps the AdS$_3$ decoupled metric in Eq.~\eqref{eq:GLMT+sw-AdS} into that of the supertube with shockwave in~\eqref{sw+twisted vacuum}; the same holds for the two-form potential. 

Therefore the natural candidate family of dual CFT states is the family obtained by fractional spectral flow with parameter $\nu=s/k$ of the family of two-charge states in Eq.~\eqref{superstube+sw CFT final}, subject to the condition of integer momentum per strand. We shall show that this condition is non-trivial, but that it is satisfied by an arbitrarily large number of states in the large $N$ limit. Recall that the lengths of the long strands $k_i$ are required to be multiples of $k$, in order that we can make this fractional spectral flow transformation.

To describe this family of states in more detail, let us introduce integer parameters $s_i$ which label the amount of spectral flow performed over the strands of length $k_i$. One has 
\begin{equation}
\nu\,=\,\frac{s}{k}\,=\,\frac{s_i}{k_i}   \qquad \forall i \,.
\end{equation}

Our proposed dual CFT states of the bulk solutions involving a shockwave on a fractionally spectral flowed supertube background in Eq.~\eqref{GLMT+sw Q1Q5} are the following states (and their superpositions):
\begin{align}
\begin{aligned}
\label{GLMT+sw CFT}
    &\ket{++}_{k,s}^{N_0}
    \ket{\varepsilon_1\bar\varepsilon_1}_{k_1,s_1}^{d_1}
    \cdots
    \ket{\varepsilon_{n_s}\bar\varepsilon_{n_s}}_{k_{n_s},s_{n_s}}^{d_{n_s}}\,,\qquad~ \frac{k_i}{k} \in \mathbb{Z},
    \qquad~ \frac{k_i}{k} \gg 1\,,
    \cr &\qquad\qquad\qquad\quad~~ k N_0\,=\,\xi N \,, \qquad \sum_{i=1}^{n_s} d_i k_i\,=\, qN\,, \qquad 
      \sum_{i=1}^{n_s} d_i \,\sim\, N^{1-\alpha} \,, \qquad 
    \alpha > 0 \,,
    \end{aligned}
    \end{align}
subject to the condition that the momentum on each CFT strand be an integer.

Let us now examine the condition of integer momentum per strand. 
For the strands corresponding to the background, recall that we have the condition $p=\frac{s(s+1)}{k}\in\mathbb{Z}$, Eq.~\eqref{eq:s-k-cond-GLMT}.
Similarly, for the strands corresponding to the shockwave, we require 
\begin{equation}
 p_i\,=\,\frac{s_i(s_i +\varepsilon_i)}{k_i}\in\mathbb{Z}   \qquad \forall~i \,.
\end{equation}
This condition is quite non-trivial, because we have noted that the candidate dual CFT states
contain strands of parametrically large $k_i$, and because the numerator is constrained by $s_i=\frac{s\:\! k_i}{k}$.
Therefore, given an allowed pair $(s,k)$, it is important to ensure that there is a set of allowed values of $k_i$ that extend to arbitrarily large positive integers. We now prove that this is indeed the case.

Let us assume without loss of generality that $s>0$, and present the proof first for $\varepsilon_i=1$. Recall that $k>0$ by definition.
Since $s$ and $(s+1)$ share no common factors and $\frac{s(s+1)}{k}\in\mathbb{Z}$, when we decompose $k$ into its prime factors, a subset of these must divide $s$, and the rest must divide $(s+1)$. We can then write the prime factorization of $k$ in the form
\begin{equation}\label{prime decomposition-0}
k\,=\,k^{(s)}k^{(s+1)}\,=\, \prod_i n^{(s)}_i \prod_j n^{(s+1)}_j\,,\qquad
k^{(s)} \,=\,  \prod_i n^{(s)}_i \,, \qquad
k^{(s+1)} \,=\,  \prod_i n^{(s+1)}_i \,,
\end{equation}
where $n^{(s)}_i$ are primes that divide $s$, and similarly for $n^{(s+1)}_i$. Repeated primes can of course occur in this decomposition, and $n^{(s)}_i \neq n^{(s+1)}_j$ for all $i,j$.
We can then factorize $s$ and $(s+1)$ as
\begin{equation}\label{prime decomposition}
    s\,=\,\hat s \:\! k^{(s)}\,,\qquad~ s+1\,=\,\hat t  \:\! k^{(s+1)}\,,
\end{equation}
where $\hat s$ and $\hat t$ are positive integers but are not necessarily prime.
We recall that the $k_i$ are multiples of $k$, such that we can write $k_i=\hat k_i k$ for positive integers $\hat k_i$. By using $s_i=s\frac{k_i}{k}$ and the decompositions in Eqs.~\eqref{prime decomposition-0},~\eqref{prime decomposition}, we have that the momentum carried by the $i$-th type of strand is given by
\begin{equation}
    p_i\,=\,\frac{s_i(s_i+1)}{k_i}\,=\,\frac{s(s\hat k_i+1)}{k}\,=\,\frac{\hat s(s\hat k_i+1)}{k^{(s+1)}}\,.
\end{equation}
Let us define $\hat{p}_i = p_i/\hat {s}$ and show that there is an infinite sequence of $\hat{k}_i$ such that $\hat{p}_i$ is a positive integer. Rearranging, we have
\begin{equation} \label{eq:bez}
\hat{p}_i k^{(s+1)} - s \:\! \hat k_i = 1 \,.
\end{equation}
Since none of the $n^{(s+1)}_j$ divide $s$, we have $\gcd(s,k^{(s+1)})=1$. B\'ezout's identity (and the extended Euclidean algorithm) then imply that there is an infinite sequence of positive integer pairs $(\hat{k}_i,\hat p_i)$ such that \eqref{eq:bez} is satisfied, and therefore there is an infinite set of $k_i$ such that $p_i\in \mathbb{Z}$.

More generally, the right-hand side of Eq.~\eqref{eq:bez} is $\varepsilon_i$. When $\varepsilon_i =-1$, B\'ezout's identity again ensures the required infinite sequence of positive integer pairs $(\hat{k}_i,\hat p_i)$. When $\varepsilon_i =0$, one can simply take $\hat{k}_i$ to be a multiple of $k^{(s+1)}$ to obtain such an infinite sequence.

The upshot is that there is an infinite family of states of the form \eqref{GLMT+sw CFT} that obey the non-trivial condition that the momentum on each strand is an integer, including strands with arbitrarily large values of $k_i$ in the large $N$ limit.\\

Let us compute the charges of the CFT states~\eqref{GLMT+sw CFT} and compare them with the gravity result in Eq.~\eqref{ADM-glmt}. 
The scalings in the second line of Eq.~\eqref{GLMT+sw CFT} will again ensure agreement.
There is no spectral flow in the right sector, so the right charges $(\bar h,\bar m)$ are the same as those for the two-charge states dual to supertubes with shockwaves~\eqref{superstube+sw CFT final}. For the left sector, we compute the charges using Eq.~\eqref{SF h j}, and derive their large-$N$ behaviour using the second line of Eq.~\eqref{GLMT+sw CFT}. Recalling that $s_i=s\frac{k_i}{k}$ and denoting subleading terms with ellipses, we obtain
\begin{equation}\label{CFT charges GLMT+sw}
\begin{aligned}
    h&\;=\;N_0\frac{k^2+4s(s+1)}{4k}+\sum_i d_i\frac{k_i^2+4s_i(s_i+\varepsilon_i)}{4k_i}
    ~=~\frac{N}{4}+\frac{N}{k^2}s(s+\xi)+\frac{s}{k}\sum_i\varepsilon_i d_i\\
    \Rightarrow \quad h&\;=\; \frac{N}{4}+\frac{N}{k^2}s(s+\xi) + \dots\;,\\
        m&\;=\;N_0\Big(s+\frac12\Big)+\sum_i  d_i\Big(s_i+\frac{\varepsilon_i}{2}\Big)~=~\frac{s N }{k}+\xi \frac{N}{2k}+\sum_i \varepsilon_i\frac{d_i}{2}\\
        \Rightarrow \quad m&\;=\; \xi \frac{N}{2k}+\frac{s N}{k}+ \dots\;.
    \end{aligned}
\end{equation}
Comparing with the gravity charges given in Eq.~\eqref{ADM-glmt} we see that the angular momentum eigenvalue $J^3=m$ explicitly agrees. We note in passing that in Eqs.~\eqref{ADM-glmt}, \eqref{CFT charges GLMT+sw}
the $s$-dependent part of the angular momentum eigenvalue $\langle J^3\rangle=m$ does not depend on $\xi$; when $s\neq 0$, the long strands contribute a finite fraction of the angular momentum of the configuration.

\begin{figure}[t]
    \centering
    \includegraphics[width=15.5cm]{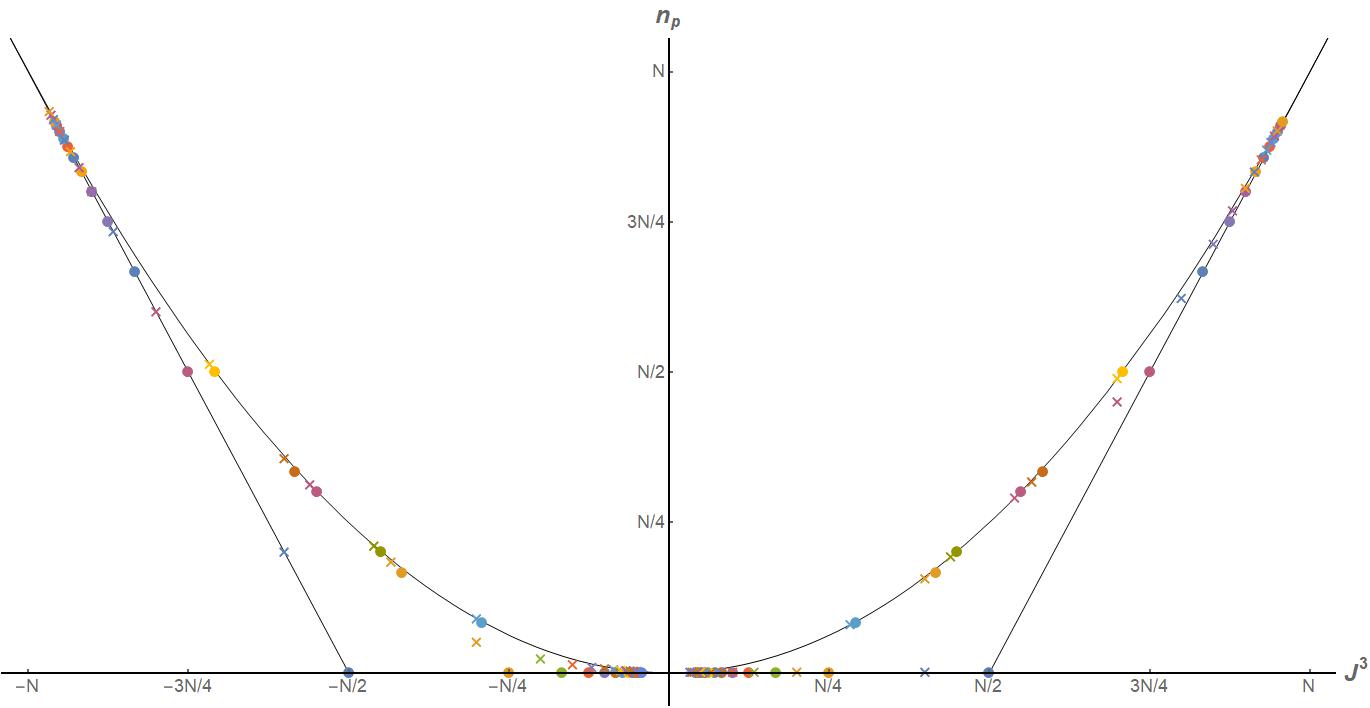}
    \caption{
Quantum numbers $(J^3,n_p)$ of spectral flowed supertubes without shockwaves (dots) and corresponding solutions with shockwaves, for $q=0.2$ (crosses). Colour coding and proximity indicate corresponding solutions. Plotted are states with $k \le 12$,  $|s|\le 12$, satisfying \eqref{eq:s-k-cond-GLMT} and with $|J^3|\le N$. All plotted points lie outside the parabola, even though some appear very close to it.}
    \label{fig:GLMT+SW}
    \vspace{2mm}
\end{figure}

To show agreement between the momentum charge $Q_p$ and the value of $h$, we extract the quantized CFT $y$-momentum charge
\begin{equation}
    n_p \,=\, h-\bar{h} \,=\, \frac{N}{k^2}s(s+\xi)\,.
\end{equation}
We then translate this CFT charge into supergravity normalization using the general relation between the supergravity charge $Q_p$ and the quantized charge $n_p$ (see e.g.~\cite[Eq.\;(6.26)]{Bena:2017xbt}), 
\begin{equation}
\label{eq:qp-comp}
    Q_p \,=\, \frac{Q_1Q_5}{R_y^2 N}n_p \,=\, \frac{Q_1Q_5}{R_y^2}\frac{s(s+\xi)}{k^2}\;.
\end{equation}
This value, derived from the CFT, is in precise agreement with Eq.~\eqref{ADM-glmt}. The agreement of conserved charges supports our proposal for the holographic description of our solutions.

In Figure~\ref{fig:GLMT+SW} we plot both the fractionally spectral flowed supertube solutions without shockwaves (dots) and the microstates obtained adding a shockwave in their core region (crosses). In all examples, the backreaction of the shockwave drives the fractionally spectral flowed supertube solutions toward the BMPV parabola, without ever reaching it except asymptotically in the limit $q \to 1$, which as we have already discussed is a singular limit. 

Let us make some observations on the set of solutions with shockwaves and their conserved charges.
First, let us consider the right-hand side of the $n_p=0$ line, i.e.~the range $0<J^3\le N/2$. Upon backreaction of the shockwave, the microstate remains on the same line. The shockwave reduces the angular momentum, corresponding to a transition from less typical to more typical two-charge microstates~\cite{Marolf:2016nwu}.

On the left-hand side of the $n_p=0$ line, when $-N/2 \le J^3<0$, the behaviour is quite different: upon adding the shockwave, the momentum charge of the microstate increases.
The difference between the two sides of the $n_p=0$ line can be understood by first noticing that in all points plotted, the shockwave adds a negative amount of $J^3$. When the background has positive $J^3$, the shockwave decreases $|J^3|$. However when the background has negative $J^3$, the shockwave increases both $|J^3|$ and the average winding of the strands in the CFT, so it is not possible for the solution with shockwave to remain on the $n_p=0$ line. A more direct understanding can be obtained by tracing the spectral flow orbits of the points on the right-hand side of the $n_p=0$ line (by fractional spectral flow with $\nu=-1/k)$.

Note that there exists a similar set of configurations with shockwaves in which the shockwave adds a positive amount of $J^3$. These can be obtained by
interchanging $\phi \leftrightarrow\psi$ in the solutions we constructed in Eq.~\eqref{GLMT+sw Q1Q5}. Their charges are obtained by reflecting Fig.~\ref{fig:GLMT+SW} in the $n_p$ axis. So in fact for each background, there exist two solutions with shockwaves of the type we have constructed, only one of which is plotted in Fig.~\ref{fig:GLMT+SW}.

The behaviour on the diagonal lines is similar to the respective halves of the two-charge line $n_p=0$. Specifically, the behaviour of the states on the left-most diagonal line is similar to that on the right-hand side of the $n_p=0$ line, being related by (integer) spectral flow with parameter $\nu = -1$. The configurations with shockwaves remain on the diagonal line. In the same way, the behaviour on the right-most diagonal line is similar to that on the left-hand half of the $n_p=0$ line. Recall that the dots on these lines include all states that have $k=1$ and all states that have $s/k \in \mathbb{Z}$ or $(s+1)/k \in \mathbb{Z}$.

The final set of dots are those that already lie close to the parabola, for which $k>1$ and neither $s/k \in \mathbb{Z}$ nor $(s+1)/k \in \mathbb{Z}$. These are the states that involve `genuinely' fractional spectral flow, in the sense that they cannot be obtained from any two-charge state by spectral flow with parameter $\nu\in\mathbb{Z}$~\cite{Giusto:2012yz}. The shockwave drives these states to be closer to the parabola, though in many cases it is not easy to see this from the plot.

We conclude this subsection by returning to the point that for the states with shockwaves that remain on the two-charge line, the process of adding a shockwave is a process that drives the system from less typical to more typical two-charge microstates~\cite{Marolf:2016nwu}. For our fractionally spectral flowed supertubes with shockwaves, making a similar interpretation is complicated by the fact that the conserved charge $n_p$ in general changes when the shockwave is added. However in both cases the solutions with shockwaves describe a family of microstates involving strands with unspecified twists $k_i$,
corresponding to the high-frequency quanta making up the shockwave that are not resolved by supergravity. 
Indeed, our proposed dual CFT states in Eq.~\eqref{GLMT+sw CFT} involve strands with windings that generically are of different lengths, including lengths scaling as positive powers of $N$. Therefore, relative to other microstates with the same respective values of $n_p$,
the states with shockwaves are naturally thought of as being more typical than the states dual to the fractionally spectral flowed supertube solutions without shockwaves.

\subsection{Interpolating between different microstates}
\label{Interpolatinggeo}

We now observe that the class of CFT states that we have studied, given in Eq.~\eqref{GLMT+sw CFT}, contains some simple examples of states that have attracted recent interest as families that interpolate between different microstate geometries~\cite{Hampton_2019,Shigemori:2021pir}. Those works studied sub-families of states of the general form
\begin{equation}\label{interpolation 0}
    \ket{++}_{k,\nu}^{N_0}\ket{++}^{d_1}_{k_1,\nu_1} \;,
\end{equation}
where the pair $(k,\nu)$ is not equal to the pair $(k_1,\nu_1)$, and coherent superpositions of such states.
We caution the reader that in this subsection we are parameterizing spectral flow with the rational parameter $\nu=s/k$ rather than the integer $s$.

This family of states is interesting because in the separate limits in which $d_1=0$ or $N_0=0$, the state reduces to a spectral flowed supertube state (or a two-charge supertube state). In~\cite{Hampton_2019} the sub-family $k=k_1=1, \nu=0, \nu_1 >0$ was studied (spectral flowed further to the NS-NS sector). 
In~\cite{Shigemori:2021pir} a general discussion was given, as well as an explicit analysis of the sub-family in which $\nu_1=\nu-\frac{1}{k_1}$. It was found that the bulk description of these states involves codimension-2 sources corresponding to an extra KKM dipole charge in the D1-D5 frame.

The family of states we have analyzed includes another distinct sub-family of states of the form~\eqref{interpolation 0}, namely that in which $k_1 \sim N^b$ with $0<b\le 1$, and either $\nu_1=\nu$ or $\nu_1=\nu+\frac{1}{k}-\frac{1}{k_1}$, i.e.
\begin{equation}\label{interpolation 0-1}
    \ket{++}_{k,\nu}^{N_0}\ket{++}^{d_1}_{k_1,\nu} ~, \qquad
    \qquad  \ket{++}_{k,\nu}^{N_0}\ket{++}^{d_1}_{k_1,\nu+\frac{1}{k}-\frac{1}{k_1}} \;.
\end{equation}
The first of these values of $\nu_1$ is obtained directly by taking the limit of our general family of CFT states~\eqref{GLMT+sw CFT} in which there is only one type of long strand, of polarization $\ket{++}$.

The second value of for $\nu_1$ arises because we have the freedom to flip the sign of the left angular momentum $J^3 $ while keeping the right angular momentum $\bar J^3$ invariant. This can be implemented by the coordinate transformation $(\psi,\phi) \to (\phi,\psi)$ in our solutions~\eqref{ADM-glmt}, as discussed below~Eq.~\eqref{eq:qp-comp}. This gives the bulk solutions dual to a set of CFT states similar to those in Eq.~\eqref{GLMT+sw CFT} but with $\ket{++}_k \to \ket{-+}_k$ and all $\varepsilon_i\to -\varepsilon_i$. This includes states of the form
\begin{equation}\label{interpolation 0b}
    \ket{-+}_{k,\nu}^{N_0}\ket{-+}^{d_1}_{k_1,\nu} \;.
\end{equation}
By shifting $\nu\to \nu+1/k$, we can rewrite these states as the second type of state in \eqref{interpolation 0-1}.

Note that setting $\nu=0$ in the second type of state in \eqref{interpolation 0-1}, we obtain a set of states of which one is a RR ground state and one is a fractional spectral flowed state,
\begin{equation}\label{interpolation 0d}
    \ket{++}_{k}^{N_0}\ket{++}^{d_1}_{k_1,\frac{1}{k}-\frac{1}{k_1}} \,.
\end{equation}
In our setup, the bulk configurations with $q=0$ correspond to CFT states with all strands of one type (of the shorter winding $k$). Dialling $q$ larger, we obtain solutions that describe states of the form~\eqref{interpolation 0} with $k_1 \sim N^b$ and either $\nu_1=\nu$ or $\nu_1=\nu+\frac{1}{k}-\frac{1}{k_1}$. For these states, $q$ controls the fraction of the total strand budget accounted for by the long strands, as discussed around Eq.~\eqref{N0 N relation}.

In the analysis of~\cite{Shigemori:2021pir}, emphasis was placed on the ability to interpolate from states involving strands of all one type to states involving strands of all the other type. On this point, let us note that there are two limitations to our construction: first, we cannot interpolate all the way to having only long strands, as this would invalidate the shockwave approximation we have made; the approximation relies on both long and short strands contributing an order-one fraction of the overall strand budget. Second, our bulk solutions do not differentiate between the polarizations of the long strands, so the same bulk solutions describe interpolations between different pairs of fractionally spectral flowed states. In this sense our bulk description is more coarse-grained than that of~\cite{Shigemori:2021pir}.
Nevertheless, we have found the bulk description of interesting examples of states of the general form~\eqref{interpolation 0}, allowing us to describe a partial interpolation between strands that have different amounts of spectral flow.

\section{Discussion}
\label{Concls}

In this paper we have exhibited the first family of asymptotically flat BPS three-charge microstate geometries involving shockwaves in their cores.
Our construction is built upon solutions that describe shockwaves in global AdS$_3\times$S$^3$. We performed a spacetime fractional spectral flow transformation and then exploited the multi-center formalism of supersymmetric solutions to construct asymptotically flat (specifically asymptotically~$\mathbb{R}^{4,1}\times$S$^{1}$) solutions. The resulting solutions are recorded in Eq.~\eqref{GLMT+sw Q1Q5}.

The solutions contain a shockwave  singularity. Away from the shockwave locus, the solutions are regular up to possible orbifold singularities that are physical in string theory. We have excluded closed timelike curves analytically in the decoupling limit, and numerically in the full asymptotically flat solutions.

We have proposed the holographic description of these supergravity solutions as being the family of 
CFT states described in Eq.~\eqref{GLMT+sw CFT}, subject to the constraint of integer momentum per CFT strand.
We observed that this constraint is non-trivial, and proved that it is satisfied by an infinite sequence of states involving strands of arbitrarily long length at large $N$.

We provided supporting evidence for our proposal by comparing conserved charges, finding precise agreement. We also performed a precision holographic test using the recently developed explicit dictionary of~\cite{Giusto:2015dfa,Rawash:2021pik}, again with exact agreement.
As usual, such tests cannot prove that the identification of the CFT dual states is precisely correct (there can be many states with same expectation value of a set of light operators), however their agreement together with the method of spectral flow used in the supergravity construction provide strong supporting evidence of this proposal for the dual CFT states.

Our solutions describe the backreaction of highly energetic supergravity quanta on a fractionally spectral flowed supertube background. While the total energy of the shockwave is fixed, the supergravity solutions do not contain information about the details of these supergravity quanta. In the holographically dual CFT, the corresponding statement is that the fraction of the total strand budget taken up by long strands in the states~\eqref{GLMT+sw CFT} is fixed, however the length, degeneracy and polarization of each long strand are not. In this sense the shockwave provides a coarse-grained description of the backreacted high-energy quanta.

We observed that the CFT states described by our solutions contain examples of states that interpolate between certain types of different three-charge microstates that have recently attracted attention, expanding upon the classes of states discussed in~\cite{Hampton_2019,Shigemori:2021pir}.

We also observed that in our asymptotically flat solutions, the location of the shockwave is not the same as the evanescent ergosurface. As a result, the addition of the shockwave does not come at zero cost in energy, and instead changes the momentum charge $n_p$ along the $y$-circle. This is a physical difference from the two-charge solutions of~\cite{Lunin:2002bj}, discussed in~\cite{Marolf:2016nwu} in the context of an evolution from less typical to more typical states following the perturbation process described in~\cite{Eperon:2016cdd}. Nevertheless, we have argued that the CFT states dual our solutions with shockwaves are naturally thought of as more typical than the solutions without shockwaves, when each is compared to other microstates with the same respective values of $n_p$.

Our results offer possibilities for generalization. By considering more general seed solutions, one could construct more general families of microstates involving shockwaves. Within such families, it may be possible to construct three-charge solutions in which the shockwave is located at the evanescent ergosurface, thus preserving the total energy and hence the value of $n_p$. Such solutions would connect more directly to the work of~\cite{Eperon:2016cdd,Marolf:2016nwu}.

It would be interesting to construct solutions involving shockwaves in the non-BPS microstate geometries of~\cite{Jejjala:2005yu}. In particular, a solution with a shockwave in the ergoregion of these backgrounds could describe the backreaction of the quanta generated by ergoregion emission, which has been interpreted microscopically as an enhanced unitary version of Hawking radiation for such microstates~\cite{Chowdhury:2007jx,Avery:2009tu}. One could further generalize this line of enquiry to more general non-BPS microstate geometries such as those of~\cite{Bena:2015drs,Bena:2016dbw,Bossard:2017vii} and~\cite{Bah:2020ogh,Bah:2020pdz,Bah:2021owp,Bah:2021rki,Heidmann:2021cms}.

Finally, two-charge solutions involving shockwaves can be obtained as limits of the general family of two-charge solutions. Such a general bulk description is not known for three-charge microstates. Our solutions may be useful data points to inform the program to construct a complete description of  general three-charge black hole microstates.

\vspace{5mm}
\section*{Acknowledgements}
We thank Davide Bufalini, Nicolas Kovensky, Emil Martinec, Stefano Massai, Michele Santagata, Masaki Shigemori, Marika Taylor and Amitabh Virmani for fruitful discussions. The work of BC was supported by STFC grant ST/T000775/1. The work of SR was supported by a Royal Society URF Enhancement Award.
The work of DT was supported by a Royal Society Tata University Research Fellowship.


\appendix

\section{Type IIB supergravity ansatz and BPS equations}\label{app:sugra}
In this appendix, we record the class of solutions to type IIB supergravity  compactified on $T^4$ within which we work. The ansatz allows for 1/8-BPS solutions with D1-D5-P charges, and in six dimensions corresponds to minimal 6D supergravity coupled to one tensor multiplet.

The ansatz is arranged as a fibration over a four-dimensional spatial base $\cB$.
Denoting by $d \hat{s}^2_4$ and $\widehat{\mathrm{vol}}_{4}$ the flat metric and the volume form on $T^4$ respectively,
the ansatz for the supergravity fields is 
\begin{align}\label{ansatzSummary}
d s^2_{10} &~=~  \,ds^2_6 +\sqrt{\frac{Z_1}{Z_2}}\,d \hat{s}^2_{4}\, ,\nonumber\\
d s^2_{6} &~=~-\frac{2}{\sqrt{\mathcal{P}}}\,(d v+\beta)\,\Big[d u+\omega + \frac{\mathcal{F}}{2}(d v+\beta)\Big]+\sqrt{\mathcal{P}}\,d s^2_4(\cB)\,,\nonumber\\
e^{2\Phi}&~=~\frac{Z_1^2}{\mathcal{P}}\, ,\qquad \qquad  \mathcal{P}   \;=\;     Z_1  Z_2\,, \\
C_2 &~=~ -\frac{Z_2}{\mathcal{P}}\,(d u+\omega) \wedge(d v+\beta)+ a^1 \wedge  (d v+\beta) + \hat\gamma_2\,,\nonumber\\ 
C_6 &~=~\widehat{\mathrm{vol}}_{4} \wedge \left[ -\frac{Z_1}{\mathcal{P}}\,(d u+\omega) \wedge(d v+\beta)+ a^2 \wedge  (d v+\beta) + \hat\gamma_1\right]\nonumber , 
%
\end{align}
where $Z_1, Z_2, \cal{F}$ are scalars, $\,\beta,\omega,a^1,a^2$ are one-forms on $\cB$,  $\,\hat\gamma_1,\hat\gamma_2$ are two-forms on $\cB$. We work in conventions in which the coordinates $u$, $v$ are related to the canonical asymptotic time $t$ and common D1-D5 spatial direction $y$ by
\begin{equation}
  u\,=\,t\,,\qquad v\,=\,t-y\,.
\end{equation}
Following~\cite{Gutowski:2003rg}, we introduce the operator
\begin{equation}
\mathcal{D} \;\equiv\; \tilde d - \beta\wedge \frac{\partial}{\partial v}\, ,
\end{equation}
where $\tilde d$ is the exterior differential on the spatial base $\cB$.

The structure of the BPS equations for this ansatz is as follows. The base metric $ds^2_4(\cB)$ and the one-form $\beta$ satisfy non-linear equations known as the ``zeroth layer''. Having solved these initial equations, the remaining BPS equations are organized into two further layers of linear equations to be solved~\cite{Giusto:2013rxa,Bena:2011dd}.

In this work we construct solutions in which the four-dimensional base space $\cB$ is flat $\mathbb{R}^4$, and in which $\beta$ is independent of $v$.
The BPS equation for $\beta$ is then
 \begin{equation}\label{eqbeta}
\tilde{d} \beta \;=\; *_4 \tilde{d}\beta\,,
 \end{equation}
where $*_4$ stands for the flat $\mathbb{R}^4$ Hodge dual. 

 We introduce the $SO(1,1)$ Minkowski metric $\eta_{ab}$ ($a=1,2$) in the form
\be \label{eq:etaab}
\eta_{12} ~=~ \eta_{21} ~=~ 1 \,.
\ee
This metric is used to raise and lower $a,b$ indices. 
We then have
\be
\mathcal{P} ~\equiv~ \frac{1}{2} \eta^{ab} Z_a Z_b \;=\; Z_1 Z_2\,.
\ee
We introduce the two-forms $\Theta^1$, $\Theta^2$ as follows:
\begin{equation}
\label{Thetadefs}
\Theta^b ~\equiv~ \mathcal{D} a^b + \eta^{bc} \:\! \dot{\hat\gamma}_c  \;.
\end{equation}
The BPS ansatz for the flux $G^1=dC_2$ is
\begin{equation}
    G^1=d\Big [-\frac{Z_2}{\mathcal{P}}\,(d u+\omega) \wedge(d v+\beta)\Big]+\star_4\mathcal{D}Z_2+(dv+\beta)\wedge\Theta^1\,.
\end{equation}
The ``first layer'' of the BPS equations is
\bea \label{eq:firstlayer}  
 *_4 \mathcal{D}\dot{Z}_a ~=~ & \eta_{ab} \mathcal{D}\Theta^{b}\,,\qquad \mathcal{D}*_4 \mathcal{D}Z_a ~=~  - \eta_{ab} \Theta^{b} \! \wedge d\beta\,,
\qquad \Theta^{a} ~=~ *_4 \Theta^{a} \,.
\eea
The ``second layer'' of the BPS equations is given by
\begin{equation}
 \begin{aligned}
\mathcal{D} \omega + *_4 \mathcal{D}\omega + \mathcal{F} \,d\beta 
~=~ & Z_a \Theta^{a}\,,  \\ 
 *_4\mathcal{D}*_4\!\Bigl(\dot{\omega} -\frac{1}{2}\,\mathcal{D}\mathcal{F}\Bigr) 
~=~& \ddot{\mathcal{P}}  -\frac{1}{2} \eta^{ab} \dot{Z}_a \dot{Z}_b 
-\frac{1}{4} \eta_{ab} *_4\! \Theta^{a}\wedge \Theta^{b} \,.
\end{aligned}
\label{eqFomega-app}
\end{equation} 

\section{Conserved charges of three-charge solutions with shockwaves} \label{charges11}

In this appendix we compute the five-dimensional conserved ADM mass and angular momenta carried by our three-charge microstate solutions with shockwaves, given in Eq.~\eqref{GLMT+sw Q1Q5}. The asymptotic metric to leading order has sphere radius $\bar{r}=\sqrt{\xi} r$ in the presence of the shockwave. With this in mind, we use  \cite[Eqs.~(2.17), (2.18)]{Harmark:2003dg} (see also \cite[Eqs.~(3.3),~(3.5)]{Harmark:2004ch}) to calculate the ADM mass of the solution in~\eqref{GLMT+sw Q1Q5},
\begin{align}
M_{ADM}&\;=\;\frac{\Omega_3 L}{16 \pi G_6} (3 c_t-c_y)\\
&\;=\;\frac{\pi }{4 G_5} \left(Q_1+Q_5+\frac{Q_1 Q_5}{R_y^2}\frac{ s (s+\xi)}{k^2}\right)\,,
\label{massADM}
\end{align} where $G_6= L \:\! G_5$, $\,L=2 \pi R_y,\,$ and $\Omega_{3}=2 \pi^2$ is the area of the unit sphere $S^{3},\,$ and where we have used $a=\frac{\sqrt{Q_1 Q_5}}{R_y}$. 
\par
To calculate the conserved five-dimensional ADM angular momenta, we dimensionally reduce on the $y$-circle.
Following the discussion around \cite[Eq.~(1.58)--(1.65)]{Myers:2011yc} and again using the coordinate $\bar{r}$, we compute the angular momentum along $\psi$, finding
\begin{align}
J^{\psi}&\;=\;-\frac{\pi }{4 G_5}\frac{a s \sqrt{Q_1 Q_5}}{ k }
\;=\;-\frac{ s \:\! n_1 n_5}{ k }\,,
\end{align} where in the second equality we have used $a=\frac{\sqrt{Q_1 Q_5}}{R_y}$, $G_5=\frac{G_{10}}{2 \pi R_y (2 \pi)^4 V_{4}}$,  
$G_{10}=8 \pi^6 g_s^2 l_s^8$, $Q_1=\frac{g_s n_1 {\alpha}^{\prime 3}}{V_{4}}$ and $Q_5=g_s n_5 {\alpha}^{\prime}$.
Similarly the angular momentum along $\phi$ is 
\begin{align}
J^{\phi}&\;=\;\frac{\pi a (s+\xi) \sqrt{Q_1 Q_5}}{4 G_5 k }
\;=\;\frac{ (s+\xi) n_1 n_5}{ k } \,.
\end{align}
Therefore the left and right angular momenta for our new solutions are
\begin{align}
\begin{aligned}
J^3&\;=\;\frac{1}{2} (J^{\phi}-J^{\psi})
\;=\;\frac{1}{2}\frac{ \xi n_1 n_5}{ k }+\frac{ s n_1 n_5}{ k }\,,\cr
\bar J^3&\;=\;\frac{1}{2} (J^{\phi}+J^{\psi})
\;=\;\frac{1}{2}\frac{ \xi n_1 n_5}{ k }\;.
\label{ADMangmom}
\end{aligned}
\end{align}

We note that to compute the ADM mass, we could equally well have used dimensional reduction combined with~\cite[Eq.~(1.65)]{Myers:2011yc}.

\section{Precision holography}
\label{app:prec-holog}

In this appendix we record several details of our precision holographic computation of Section~\ref{sec:CFT supertube+sw}.

\subsection{Operators of the D1D5 CFT}\label{operators D1D5 CFT}

We first collect the definitions of the scalar chiral primary operators (CPOs) of scaling dimension one and two relevant for the holographic dictionary discussed in Section~\ref{sec:CFT supertube+sw}. We remind the reader that we label with $h,m$ ($\bar h,\bar m$) the left (right) conformal dimensions and R-symmetry charges of the operators.

Let us start with the CPOs of scaling dimension $\Delta=h+\bar h=1$. First, we have the currents (all sums over copy indices $r,s$ run from 1 to $N$ unless otherwise indicated): \begin{equation}\label{op J}
J^{+}\;=\;\sum_r J^{+}_{(r)}\;=\;\sum_r \psi^{+1}_{(r)}\psi^{+2}_{(r)}~,\qquad \bar J^{+}\;=\;\sum_r \bar J^{+}_{(r)}\;=\;\sum_r \bar \psi^{+1}_{(r)}\bar \psi^{+2}_{(r)} ~.
\end{equation}
Second, we have the twist-two operator $\Sigma^{++}_2$: it is composed of a `bare' twist-two operator $\sigma_{(rs)}$ associated with the permutation $(rs)$ and spin fields $S^{+},\bar S^{+}$ that map NS to R boundary conditions, and vice versa. It has dimension $(\frac{1}{2},\frac{1}{2})$ and is given by 
\begin{equation}\label{op Sigma}
\Sigma^{++}_2 \;=\; \sum_{r<s}S^{+}\bar S^{+}\sigma_{(rs)} \;=\; \sum_{r<s} \sigma_{(rs)}^{++} \,.
\end{equation}
These operators are the building blocks of the $\Delta=2$ double-trace operators that enter in the linear combination of $\tilde{\Sigma}_3$ in Eq.~\eqref{tilde Sigma_3}. Their explicit definitions are \begin{equation}\label{eq:doubletraces}
\begin{aligned}
&(\Sigma_2 \cdot \Sigma_2)^{++}\;=\; \frac{2}{N^2} \sum_{(r<s), (p<q)} \sigma_{(rs)}^{++} \sigma_{(pq)}^{++}\,,\qquad~~  (J \cdot \bar J)^{++}\;=\;\frac{1}{N} \sum_{r, s} J^{+}_{(r)} \bar J^{+}_{(s)}\,,
\end{aligned}
\end{equation}
where the numerical factors are arranged so that these operators have unit norm in the large-$N$ limit.
Both these operators are highest-weight states of $SU(2)_L\times SU(2)_R$; 
their R-symmetry descendants can be constructed by acting with the zero modes of $J^{-},\bar{J}^{-}$. 
We shall follow the conventions of~\cite{Giusto:2019qig,Rawash:2021pik} and define the descendant operators to have the same norm as the highest-weight operator.

Next, we introduce the relevant CPOs at dimension $\Delta=2$. In the untwisted sector we have the single-trace product of the holomorphic and anti-holomorphic currents:
\begin{equation}\label{op Omega}
\Omega^{++}\;=\;\sum_r J^{+}_{(r)}\bar J^{+}_{(r)}\;=\;\sum_r  \psi^{+1}_{(r)}\psi^{+2}_{(r)}  \bar \psi^{+1}_{(r)}\bar \psi^{+2}_{(r)} \,.
\end{equation}
Second, we have a twist-three operator
\begin{equation}\label{op Sigma3}
\Sigma^{++}_3\;=\;\sum_{q<r<s} \bar{J}_{-1/3}^{+}J_{-1/3}^{+}\big(\sigma_{(qrs)}+\sigma_{(qsr)}\big)\,,
\end{equation}
where $\sigma_{(qrs)},\sigma_{(qsr)}$ are bare twist operators associated with the inequivalent permutations $(qrs)$ and $(qsr)$; the current mode insertions add the necessary charge to obtain a chiral primary.

\subsection{Precision holographic test for more general states}\label{app: precision holography}

We now describe the computation of the expectation value of the single-particle operator $\tilde\Sigma_3^{00}$ on the following class of states, which is more
general than that considered in Section~\ref{sec:CFT supertube+sw}.
\begin{equation}\label{superstube+sw CFT pre app}
    \ket{++}_1^{N_0}
    \Bigg(
    \prod_{i=1}^{n_s^{+}}\ket{++}_{k_i^{+}}^{d_i^{+}} \Bigg)
    \Bigg(
    \prod_{j=1}^{n_s^{-}}\ket{--}_{k_j^{-}}^{d_j^{-}}
    \Bigg)
    \,,\qquad\quad
    N_0 +\sum_{i=1}^{n_s^{+}} d_i^{+} k_i^{+}+\sum_{i=1}^{n_s^{-}} d_i^{-} k_i^{-}=N\,.
\end{equation}
Here the superscript $\pm$ refers to the strand polarizations $\ket{++}$ and $\ket{--}$; for ease of notation we introduce the index $m=\pm$ which we shall use in some of the following expressions.

Let us first consider the contribution from $\Sigma_3^{00}$. 
Proceeding as explained after Eq.~\eqref{process Sigma_3}, and 
using Eqs.~\eqref{c++} and~\eqref{c-- c+-} from the following subsection, one obtains that the expectation value of $\Sigma_3^{00}$ on the full state~\eqref{superstube+sw CFT pre} arises from the process
\begin{equation}\label{process Sigma_3 app}
\begin{aligned}
    \Sigma_3^{00}\bigg(&\ket{++}_1^{N_0}\prod_{i=1}^{n_s^{+}}\ket{++}_{k_i^{+}}^{d_i^{+}}\prod_{j=1}^{n_s^{-}}\ket{--}_{k_j^{-}}^{d_j^{-}}\bigg)\;=\;\\&\Bigg[\sum_i\frac{(k_i^++1)^2}{6 k_i^+} N_0 d_i^++\sum_i\frac{(k_i^-)^2+6k_i^-+1}{6 k_i^-} N_0 d_i^-+
    \sum_{m,i\neq j}\frac{(k_i^m+k_j^m)^2}{6 k_i^m k_j^m} d_i^m d_j^m\\
    &\qquad{}+\sum_{i, j}\frac{(k_i^+)^2+6k_i^+k_j^-+(k_j^-)^2}{6 k_i^+ k_j^-} d_i^+ d_j^-\big(1-\delta_{k_i^+k_j^-}\big)\Bigg]\bigg(\ket{++}_1^{N_0}\prod_{i=1}^{n_s^{+}}\ket{++}_{k_i^{+}}^{d_i^{+}}\prod_{j=1}^{n_s^{-}}\ket{--}_{k_j^{-}}^{d_j^{-}}\bigg)\,,
    \end{aligned}
    \end{equation}
where the indices $i,j$ run from $1$ to $n_s^+$ ($n_s^-$) when $m=+$ ($m=-$).

Second, we consider the operator $\Omega^{00}$. By using Eq.~\eqref{eigenvalue Omega00}, it acquires a non-vanishing expectation value via the process
\begin{equation}\label{process Omega app}
    \Omega^{00}\bigg(\ket{++}_1^{N_0}\prod_{i=1}^{n_s^{+}}\ket{++}_{k_i^{+}}^{d_i^{+}}\prod_{j=1}^{n_s^{-}}\ket{--}_{k_j^{-}}^{d_j^{-}}\bigg)=\bigg(\frac{N_0}{2}+\sum_{i,m}\frac{d_i^m}{2 k_i^m} \bigg)\bigg(\ket{++}_1^{N_0}\prod_{i=1}^{n_s^{+}}\ket{++}_{k_i^{+}}^{d_i^{+}}\prod_{j=1}^{n_s^{-}}\ket{--}_{k_j^{-}}^{d_j^{-}}\bigg)\,.
\end{equation}

Third, we consider the double-trace operator $\big(J\cdot \bar J\big)^{00}$. Its expectation value arises from the process described after Eq.~\eqref{process JJ},
\begin{align}\label{process JJ app}
  & \big(J\cdot \bar J\big)^{00}\bigg(\ket{++}_1^{N_0}
  \prod_{i=1}^{n_s^{+}}\ket{++}_{k_i^{+}}^{d_i^{+}}\prod_{j=1}^{n_s^{-}}\ket{--}_{k_j^{-}}^{d_j^{-}}\bigg)\;=\;\\
&   \Bigg[\frac{N_0^2}{2}+ N_0\sum_i d_i^+- N_0\sum_i d_i^-+
  \!
   \sum_{m,i,j}\frac{d_i^m d_j^m}{2}-\sum_{i, j}d_i^+d_j^- \Bigg]\bigg(\ket{++}_1^{N_0}\prod_{i=1}^{n_s^{+}}\ket{++}_{k_i^{+}}^{d_i^{+}}\prod_{j=1}^{n_s^{-}}\ket{--}_{k_j^{-}}^{d_j^{-}}\bigg)\,.\nonumber
\end{align}

By combining the definition of the single-particle operator $\tilde\Sigma_3^{00}$ with Eqs.~\eqref{process Sigma_3 app}--\eqref{process JJ app}, we obtain the expectation value of the single-particle operator. We find cancellation of all terms that are clearly of order $N^{1/2}$, leaving the following remainder: 
\begin{equation}\label{VeV tilde Sigma3-app}
\begin{aligned}
    \!\!\big\langle\tilde\Sigma_3^{00}\big\rangle\;=\;\frac{1}{N^{3/2}}\Bigg[& N_0\sum_{m,i} d_i^m +\sum_{m,i\neq j}d_i^m d_j^m \frac{(k_j^m)^2+3k_i^m k_j^m }{4k_i^m k_j^m}
    +\sum_{i, j}d_i^+d_j^- 
    \Big(1-2\delta_{k_i^+,k_j^-}\Big)
    \Bigg]\,.
    \end{aligned}
\end{equation}
We must ensure that this remainder is 
subleading compared to $N^{1/2}$.
When the long strands were all of polarization $\ket{++}$, this condition led to the constraint $\sum_id_i \sim N^{1-\alpha}$ with $\alpha>0$. We will obtain the analogous constraint, however to do so we must take care since now~\eqref{VeV tilde Sigma3-app} is not the sum of positive terms, due to the final term.

Let us therefore examine the final term. Without loss of generality, let us assume $n_s^+ \ge n_s^-$. 
To obtain a bound on this term, let us consider the worst-case scenario in which $k_i^+=k_i^-$ for all $i=1,\ldots,n_s^-$.
The magnitude of the negative contribution to this term is then given by
\begin{equation} \label{eq:bound-app}
    \frac{1}{N^{3/2}}\sum_{i=1}^{n_s^-} d_i^+d_i^- \,.
\end{equation}
Since no $d_i^m$ can scale as $N$, and since $N_0\sim N$, the magnitude of this term is subleading with respect to the first term in \eqref{VeV tilde Sigma3-app}. 
Therefore these terms cannot cancel each other, and so the first term in \eqref{VeV tilde Sigma3-app} must by itself be subleading with respect to $N^{1/2}$. This implies that:
\begin{equation} \label{eq:cond-pm}
    \left(\sum_{i=1}^{n_s^+} d_i^++\sum_{i}^{n_s^-} d_i^-\right)\;\sim\; N^{1-\alpha}\,,\qquad\ \alpha>0 \;.
\end{equation}
Upon imposing this condition, the other terms in~\eqref{VeV tilde Sigma3-app} are also subleading with respect to $N^{1/2}$, using similar reasoning to that used in the main text. We thus find that the condition \eqref{eq:cond-pm} is necessary and sufficient for the precision holographic test to be passed for this more general class of states. The completely general case is analogous.

\subsection{Fusion coefficients for $\Sigma_3$} \label{app:fusion sigma3}

In this final subsection we compute the fusion coefficients $c_{k_1k_2}$ for the following processes:
\begin{equation}\label{ck1k2 def}
\begin{aligned}
    \sigma_3^{00}\ket{++}_{k_1}\ket{++}_{k_2}&\;=\;c_{ k_1k_2}^{(++)}\big(1-\delta_{k_1,k_2}\big)\ket{++}_{k_1}\ket{++}_{k_2}\,,\\
    \sigma_3^{00}\ket{--}_{k_1}\ket{--}_{k_2}&\;=\;c_{ k_1k_2}^{(--)}\big(1-\delta_{k_1,k_2}\big)\ket{--}_{k_1}\ket{--}_{k_2}\,,\\
    \sigma_3^{00}\ket{++}_{k_1}\ket{--}_{k_2}&\;=\;c_{ k_1k_2}^{(+-)}\big(1-\delta_{k_1,k_2}\big)\ket{++}_{k_1}\ket{--}_{k_2}\,.
    \end{aligned}
\end{equation}
The factor $(1-\delta_{k_1,k_2})$ can be explained as follows. The operator $\sigma_3^{00}$ corresponds to a three-cycle that, when acting on two permutations of length $k_1$ and $k_2$, produces another pair of permutations of length $k_1$ and $k_2$ by shuffling the copies~\cite{Giusto:2019qig}. This process can occur only if $k_1\neq k_2$. 

We now give an explicit derivation of the coefficient $c_{ k_1k_2}^{(++)}$. The derivation of the coefficients $c_{ k_1k_2}^{(--)}$ and $c_{ k_1k_2}^{(+-)}$ is analogous, and we simply report their values at the end of the appendix.

We compute the coefficient $c^{(++)}_{k_1k_2}$ by requiring that the precision holography dictionary~\eqref{holographic dictioanry Sigma3} for the single-particle operator $\tilde{\Sigma}_3$ holds on the two-charge CFT state:
\begin{equation}\label{CFT state fusion}
    \sum_{N_1} \Big( A\ket{++}_{k_1}\Big)^{N_1}\Big(B\ket{++}_{k_2} \Big)^{N-N_1}\,,
\end{equation}
where, for concreteness, we take $k_1\neq k_2$.
Here $A,B$ are coefficients that we take to be real; they are related to the average number of strands $\bar N_1$ and $N-\bar N_1$ via~\cite{Giusto:2015dfa}
\begin{equation}\label{average number strand}
    k_1 \bar N_1=A^2\,, \qquad k_2 (N-\bar N_1)=B^2\,.
\end{equation}

Let us first compute the bulk quantity $\Big[s^{(6)(-a,-\dot a)}_{k=2}\Big]$ defined in Eq.~\eqref{s6k2}. We do so by generating the harmonic functions $Z_1$ and $Z_2$ as in \cite[Eq. (B.2)]{Giusto:2015dfa}, making use of the following profile functions:
\begin{equation}
    g_1(v')\;=\;\frac{a}{k_1}e^{\frac{2\pi i k_1}{L}v'}+\frac{b}{k_2}e^{\frac{2\pi i k_2}{L}v'}\,,\qquad g_{i\neq1}(v')\;=\;0\,.
\end{equation}
The supergravity Fourier modes $a,b$ are related to the CFT coefficients $A,B$ via
\begin{equation}\label{A to a}
A\;=\;R_y\sqrt{\frac{N}{Q_1Q_5}}a\,,\qquad B\;=\;R_y\sqrt{\frac{N}{Q_1Q_5}}b\,,
\end{equation}
and satisfy the relation
\begin{equation}\label{constraint a b}
    a^2+b^2\;=\;\frac{Q_1Q_5}{R_y^2}\,.
\end{equation}
Upon performing the asymptotic expansion in Eq.~\eqref{eq:geometryexpansion}, one finds that the spin component $(0,0)$ is non-vanishing, with value
\begin{equation}
    \Big[s^{(6)(0,0)}_{\k=2}\Big]\;=\;\sqrt{2}\frac{a^2b^2}{k_1k_2}\frac{R_y^4}{(Q_1Q_5)^2}\,.
\end{equation}

The holographic dictionary in Eq.~\eqref{holographic dictioanry Sigma3} then predicts that the single-particle scalar CPO $\tilde{\Sigma}_3^{00}$ has the following expectation value on the CFT state~\eqref{CFT state fusion}:
\begin{equation}\label{VeV tilde sigma fusion}
    \big\langle\tilde{\Sigma}_3^{00}\big\rangle\;=\;\frac{a^2b^2}{k_1k_2}\frac{R_y^4}{(Q_1Q_5)^2}N^{1/2}\,.
\end{equation}
We now fix the fusion coefficient $c_{k_1k_2}$ by requiring that this is indeed the case. 
The CFT operators in the linear combination~\eqref{tilde Sigma_3} 
that contribute at leading order at large $N$ to the expectation value of the single-particle operator $\tilde{\Sigma}^{00}_3$ are the single-trace operators $\Sigma_3^{00}$ and $\Omega^{00}$ and the double-trace $\big(J\cdot \bar J\big)^{00}$.

First, we consider the operator $\Sigma_3^{00}$. Its expectation value is obtained by multiplying the fundamental process~\eqref{ck1k2 def} by the number of different ways the twist operator can act on the coherent state, as we shall describe momentarily. When the operator $\Sigma_3^{00}$ acts on a term in the coherent state sum~\eqref{CFT state fusion}, the contribution is
\begin{equation}
    \Sigma_3^{00}\Big[ \ket{++}_{k_1}^{N_1}\ket{++}_{k_2}^{N-N_1}\Big]\;=\;c_{k_1k_2}^{(++)} N_1(N-N_1)k_1k_2\Big[ \ket{++}_{k_1}^{N_1}\ket{++}_{k_2}^{N-N_1}\Big]\,.
\end{equation}
The numerical factor $N_1(N-N_1)$ follows from the fact that the twist operator can act on any of the $N_1(N-N_1)$ pairs of $\ket{++}_{k_1}$, $\ket{++}_{k_2}$, while the term $k_1 k_2$ occurs because each strand can be cut in $k_1$ and $k_2$ different positions respectively.
Using Eqs.~\eqref{average number strand} and~\eqref{A to a} we find
\begin{equation}\label{VeV sigma3 fusion}
\big\langle \Sigma_3^{00}\big\rangle\;=\;c_{k_1k_2}^{(++)} \,a^2\, b^2\frac{R_y^4 N^2}{(Q_1Q_5)^2}\,.
\end{equation}

Second, we consider the operator $\Omega^{00}$.
The relevant contribution to the expectation value of $\Omega^{00}$ then follows from Eq.~\eqref{eigenvalue Omega00} via the basic process
\begin{equation}
    \Omega^{00}\Big[ \ket{++}_{k_1}^{N_1}\ket{++}_{k_2}^{N-N_1}\Big]\;=\;\Big(\frac{N_1}{2k_1}+\frac{N-N_1}{2k_2}\Big)\Big[ \ket{++}_{k_1}^{N_1}\ket{++}_{k_2}^{N-N_1}\Big]\,.
\end{equation}
It follows from Eqs.~\eqref{average number strand}--\eqref{constraint a b} that
\begin{equation}
    \big\langle\Omega^{00}\big\rangle\;=\;(a^2+b^2)\left(\frac{a^2}{2k_1^2}+\frac{b^2}{2k_2^2}\right)\frac{R_y^4 N}{(Q_1Q_5)^2}\,.
\end{equation}

Third, we consider the double-trace operator $\big(J\cdot \bar J\big)^{00}=\frac{2}{N}\sum_{r,s}J_{(r)}^{3}\bar J_{(s)}^{3}$. When acting on a member of the coherent state~\eqref{CFT state fusion}, this operator produces three terms, which correspond to: (i) both left and right currents acting on a strand of twist $k_1$, (ii) both currents acting on a strand of length $k_2$, and (iii) each current acting on a different type of strand. This produces the following contribution:
\begin{equation}
    \big(J\cdot \bar J\big)^{00}\Big[ \ket{++}_{k_1}^{N_1}\ket{++}_{k_2}^{N-N_1}\Big]\;=\;\frac{2}{N}\Big(\frac{N_1^2}{4}+\frac{N_1(N-N_1)}{2}+\frac{(N-N_1)^2}{4}\Big)\Big[ \ket{++}_{k_1}^{N_1}\ket{++}_{k_2}^{N-N_1}\Big]\,,
\end{equation}
which implies
\begin{equation}
    \big\langle\big(J\cdot \bar J\big)^{00}\big\rangle\;=\;\Big(\frac{a^4}{2k_1^2}+\frac{a^2 b^2}{k_1k_2}+\frac{b^4}{2k_2^2}\Big)\frac{R_y^4 N}{(Q_1 Q_5)^2}\,,
\end{equation}
where we have used Eqs.~\eqref{average number strand} and~\eqref{A to a}.
By using the definition of the single-particle operator $\tilde{\Sigma}_3$ in Eq.~\eqref{tilde Sigma_3}, we have that the holographic prediction in Eq.~\eqref{VeV tilde sigma fusion} holds provided that
\begin{equation}\label{c++}
    c_{k_1k_2}^{(++)}\;=\;\frac{(k_1+k_2)^2}{6k_1^2k_2^2}\,.
\end{equation}
With similar computations, one obtains
\begin{equation}\label{c-- c+-}
    c_{k_1k_2}^{(--)}\;=\;\frac{(k_1+k_2)^2}{6k_1^2k_2^2}\,,\qquad c_{k_1k_2}^{(+-)}\;=\;\frac{k_1^2+6k_1 k_2+k_2^2}{6k_1^2k_2^2} \,.
\end{equation}

\newpage

\bibliographystyle{utphys}      
\bibliography{microstates}     


\end{document}